\documentclass[10pt]{iopart}
\pdfoutput=1

\usepackage{graphicx}
\usepackage{tabularx}
\usepackage{multirow}
\usepackage{color}
\usepackage{cite}
\usepackage{iopams}  
\usepackage[hyphens]{url}  
\usepackage{hyperref}
\usepackage[T1]{fontenc}

\begin{document}

\title{Direct comparisons of European primary and secondary frequency standards via satellite techniques} 
\author{F. Riedel$^1$, A. Al-Masoudi$^1$, E. Benkler$^1$, S. D\"orscher$^1$, V. Gerginov$^1$, C. Grebing$^1$, S. H\"afner$^1$, N. Huntemann$^1$, B. Lipphardt$^1$, C. Lisdat$^1$, E. Peik$^1$, D. Piester$^1$, C. Sanner$^1$, C. Tamm$^1$, S. Weyers$^1$, H. Denker$^2$, L. Timmen$^2$, C. Voigt$^2$, D. Calonico$^3$, G. Cerretto$^3$, G. A. Costanzo$^{3,4}$, F. Levi$^3$, I. Sesia$^3$, J. Achkar$^5$, J. Gu\'ena$^5$, M. Abgrall$^5$, D. Rovera$^5$, B. Chupin$^5$, C. Shi$^5$, S. Bilicki$^5$, E. Bookjans$^5$, J. Lodewyck$^5$, R. Le Targat$^5$, P. Delva$^5$, S. Bize$^5$, F. N. Baynes$^6$, C. F. A. Baynham$^6$, W. Bowden$^6$, P. Gill$^6$, R. M. Godun$^6$, I. R. Hill$^6$, R. Hobson$^6$, J. M. Jones$^6$, S. A. King$^6$, P. B. R. Nisbet-Jones$^6$, A. Rolland$^6$, S. L. Shemar$^6$, P. B. Whibberley$^6$ and H. S. Margolis$^6$ }

\address{$^1$Physikalisch-Technische Bundesanstalt, Bundesallee 100, 38116 Braunschweig, Germany}
\address{$^2$Institut f\"ur Erdmessung, Leibniz Universit\"at Hannover, 30167 Hannover, Germany}
\address{$^3$Istituto Nazionale di Ricerca Metrologica INRIM, Torino, Italy}
\address{$^4$Dipartimento di Elettronica e Telecomunicazioni, Politecnico di Torino, Torino, Italy }
\address{$^5$LNE-SYRTE, Observatoire de Paris, Universit\'e PSL, CNRS, Sorbonne Universit\'e, 61 avenue de l'Observatoire, 75014 Paris, France}
\address{$^6$National Physical Laboratory, Teddington, UK}

\ead{erik.benkler@ptb.de}

\vspace{10pt}
\begin{indented}
\item[]\today
\end{indented}

\begin{abstract}
We carried out a 26-day comparison of five simultaneously operated optical clocks and six  atomic fountain clocks located at INRIM, LNE-SYRTE, NPL and PTB by using two satellite-based frequency comparison techniques: broadband Two-Way Satellite Time and Frequency Transfer (TWSTFT) and Global Positioning System Precise Point Positioning (GPS PPP). With an enhanced statistical analysis procedure taking into account correlations and gaps in the measurement data, combined overall uncertainties in the range of $1.8 \times 10^{-16}$ to $3.5 \times 10^{-16}$ for the optical clock comparisons were found. 
The comparison of the fountain clocks yields results with a maximum relative frequency difference of $6.9 \times 10^{-16}$, and combined overall uncertainties in the range of $4.8 \times 10^{-16}$ to $7.7 \times 10^{-16}$.
\end{abstract}

%
%
%
%
\ioptwocol

\section{Introduction}\label{sec:intro}

Primary and secondary frequency standards find broad application in metrology, e.g. for steering time scales, and in fundamental research. Currently, the definition of the SI second is based on the ground-state hyperfine transition in the caesium-133 atom, with fountain clocks providing the best realizations with respect to accuracy and frequency instability, reaching relative uncertainties in the low 10$^{-16}$ range~\cite{Wynands2005, Levi2014, Guena2012a, wey18}. Optical frequency standards, on the other hand, with relative frequency uncertainties that range down below 10$^{-18}$~\cite{nic15, Huntemann2016, gre18, bre19}, are promising candidates for a new definition of the SI second~\cite{Riehle2015, Gill2016, biz19}. 
Comparisons of frequency standards in different laboratories, both by absolute frequency measurements and by frequency ratio measurements, are necessary to provide consistency checks of the performance and accuracy of the various clocks. This is of utmost importance to ensure the continuity of time scales in the case of a redefinition. 
Some direct fountain clock comparisons have taken place in the past~\cite{Parker2001, Bauch2005, Fujieda2007, Zhang2014, Guena2017}, but only a few bilateral remote comparisons of optical clocks have been performed so far~\cite{yam11, Hachisu2014, tak16, Lisdat2016, Leute2016, gre19}.

In this work we present remote comparisons of primary and secondary frequency standards located at four European metrology laboratories: the Italian Istituto Nazionale di Ricerca Metrologica (INRIM), the French Laboratoire National de m\'{e}trologie et d'Essais - SYst\`{e}me de R\'{e}f\'{e}rences Temps-Espace (LNE-SYRTE), the UK National Physical Laboratory (NPL) and the German Physikalisch-Technische Bundesantalt (PTB). 

For this purpose we employed different satellite-based comparison techniques: Two-Way Satellite Time and Frequency Transfer (TWSTFT), which uses geostationary satellites with transmission frequencies in the Ku-band, and time and frequency transfer via satellites of Global Navigation Satellite Systems (GNSS), the Global Positioning System (GPS) in our case. In contrast to optical-fibre-based techniques, these are applicable at the intercontinental scale and it is easier to extend the link network, because only local installations are required to add an additional station. 
Since these techniques are used on a daily basis for remote comparisons of clocks currently contributing to the computation of TAI/UTC (International Atomic Time and Coordinated Universal Time), all involved laboratories already had the basic required measurement equipment and infrastructure. However, temporary setup changes and refinements were needed. 

The instability of TWSTFT is limited by the modulation rate of the pseudo-random code. Using a higher bandwidth for the satellite transmission compared to the routine TWSTFT measurements allows us to increase the modulation rate. As a consequence, the short-term instability at averaging times up to about 100 s is reduced, similar to an increase of the signal-to noise-ratio, and lower instabilities can also be expected for longer averaging times~\cite{Piester2008a}.

Here we present the results of a 26-day measurement campaign, where a simultaneous comparison of five optical clocks and six caesium  or rubidium fountain clocks, located at INRIM, LNE-SYRTE, NPL and PTB was carried out by using both broadband TWSTFT and GPS PPP.
This was the first time that the full available modulation rate of commercially available TWSTFT modems (20 Mchip/s) was used for optical clock comparisons. The campaign also represented the most comprehensive coordinated simultaneous operation of optical clocks to date. 

Section~\ref{sec:exp} describes the technical setups of all building blocks of the experiment such as the clocks, the local hydrogen masers serving as flywheel oscillators for connecting the satellite equipment to the clocks, and the satellite links. Data analysis follows in section~\ref{sec:ana}, and was a major challenge of this work, since time-deviation data from the microwave links, which was dominated by white phase noise, had to be combined with relative frequency deviaton data from the measurements between the optical clocks and the hydrogen masers, in the presence of clock and link downtimes. The non-standard analysis procedure for the comparison of the optical clocks is explained in more detail in the same section, together with a description of the data processing for the fountain clock comparisons. The results are presented and discussed in section~\ref{sec:res}, and a conclusion is given in section~\ref{sec:con}.

\section{Experimental realization}\label{sec:exp}

\begin{figure*}[]
\centering
\includegraphics[width=170mm]{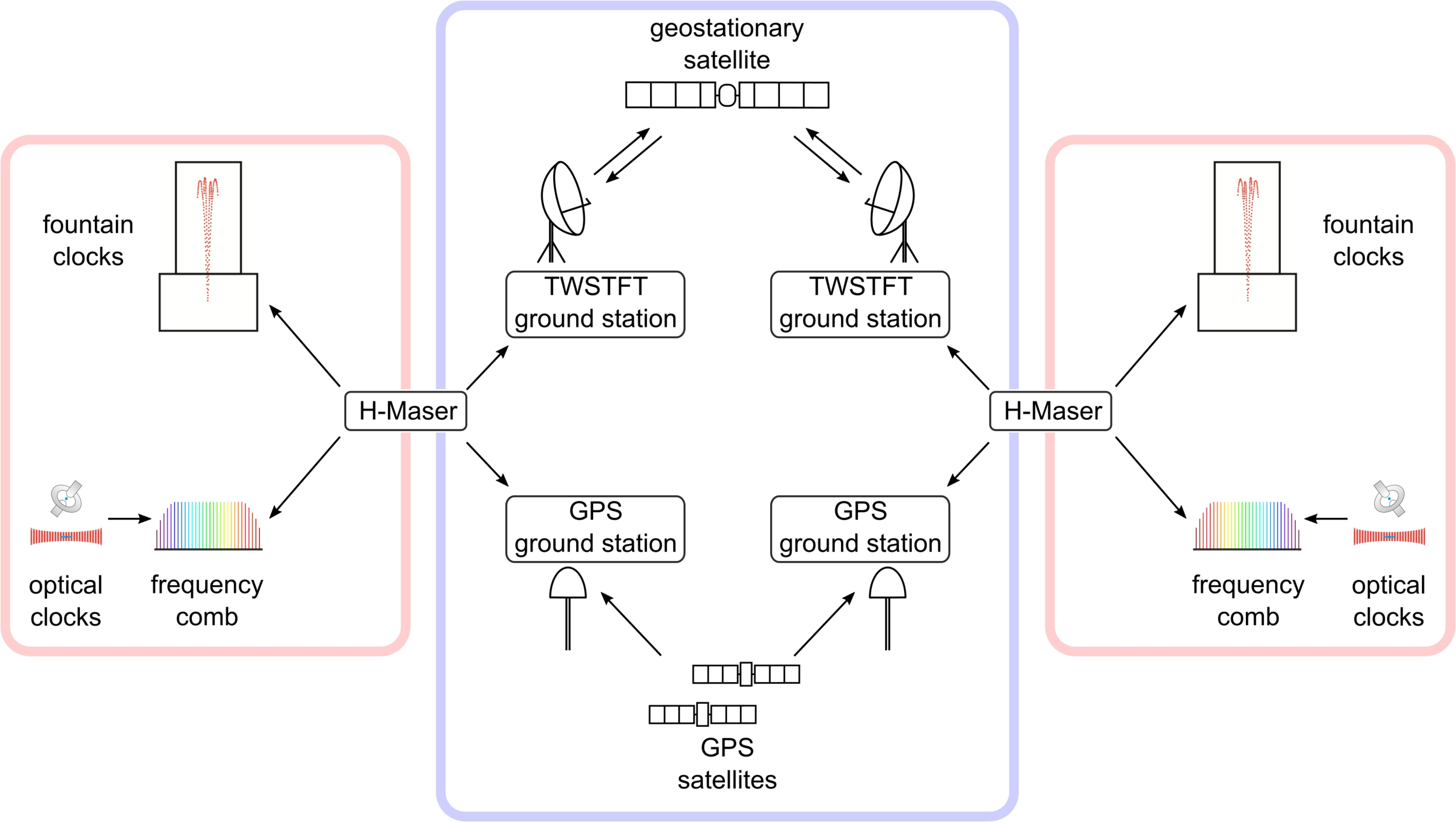}
	\caption{Basic setup of two remote stations. At each side, a continuously operating hydrogen maser was used as reference for both the satellite time transfer equipment and the atomic clocks. In this way, the clocks could be measured locally against the hydrogen masers (red boxes), which were remotely compared via the satellite link (blue box).}
	\label{fig:setup}
\end{figure*}

The setup for the remote comparison of the atomic clocks via satellite is depicted in fig.~\ref{fig:setup}. At each institute, a hydrogen maser (HM) was used as a continuously operating flywheel oscillator, simultaneously serving as the reference for the satellite time transfer equipment and being measured against the atomic clocks. As a consequence, fluctuations of the flywheel oscillator are cancelled out in the clock comparison to the extent given by common mode rejection as detailed below. While the frequency comparisons between the fountain clocks and the HMs were directly carried out in the radio frequency domain, the comparisons between the optical clocks and the HMs used optical frequency combs for the transfer between the optical and radio frequency domains. The satellite microwave links provided information on the phase and thus frequency difference between the HMs in the different institutes. In this constellation, each HM contributed to both the remote HM comparisons and the local measurements of clocks versus the HMs. In the ideal case these HM contributions would be identical and thus would cancel out completely in the remote clock comparison. In practice, the HM data were exploited over slightly different time periods in the local clock measurements and in the link comparisons in order to enable suppression of white phase noise on the satellite link data in the data processing.
This will be discussed in more detail in section~\ref{sec:ana}. 

\subsection{Clocks operated during the campaign}\label{sec:clocks}

\begin{table*}
\caption{\label{tab:clock_prop} Overview of the clocks that were compared and their estimated systematic uncertainties $u_\mathrm{B}$. In addition, the uptimes during the campaign are listed. For the optical clocks, the uptime refers to the whole duration of the campaign, whereas for the fountain clocks two different measurement intervals were analysed (see sect.~\ref{sec:ana}). This is the reason why in three cases an uptime span is indicated.} 
\begin{center}
\footnotesize
\lineup

\begin{tabular}{@{}ccccc}

\br

Institute & Clock & $u_\mathrm{B}$ & Uptime & References \\ \ms 
\mr

\multirow{1}{2.3cm}{\centering INRIM} 
& \multirow{1}{3cm}{\centering ITCsF2} 
& \multirow{1}{1.6cm}{\centering $2.3 \times 10^{-16}$} 
& \multirow{1}{1.6cm}{\centering 97$\%$} 
& \multirow{1}{2cm}{\centering \cite{Levi2014}} \\
\rule[-2ex]{0pt}{3ex}\\

\multirow{4}{2.3cm}{\centering LNE-SYRTE} 
& \multirow{4}{3cm}{\centering Sr2 $^{87}$Sr lattice\linebreak FO1\linebreak FO2\linebreak FO2-Rb} 
& \multirow{4}{1.6cm}{\centering $4.1 \times 10^{-17}$\linebreak $3.6 \times 10^{-16}$\linebreak $2.5 \times 10^{-16}$\linebreak $2.7 \times 10^{-16}$} 
& \multirow{4}{1.6cm}{\centering 68$\%$\linebreak 71$\%$ -- 75$\%$\linebreak 77$\%$ -- 79$\%$\linebreak 80$\%$}	
& \multirow{4}{2cm}{\centering \cite{Lisdat2016, lod16}\linebreak \cite{Guena2012a}\linebreak \cite{Guena2012a, gue11}\linebreak \cite{Guena2012a, Guena2014}} \\
\rule[-2ex]{0pt}{11ex}\\

\multirow{2}{2.3cm}{\centering NPL} 
& \multirow{2}{3cm}{\centering $^{87}$Sr lattice\linebreak $^{171}$Yb$^+$ single-ion (E3)} 
& \multirow{2}{1.6cm}{\centering $6.8 \times 10^{-17}$\linebreak $1.1 \times 10^{-16}$} 
& \multirow{2}{1.6cm}{\centering 77$\%$\linebreak 74$\%$}	
& \multirow{2}{2cm}{\centering \linebreak \cite{Baynham2018} } \\
\rule[-2ex]{0pt}{6ex}\\

\multirow{4}{2.3cm}{\centering PTB} 
& \multirow{4}{3cm}{\centering $^{87}$Sr lattice\linebreak $^{171}$Yb$^+$ single-ion (E3)\linebreak CSF1\linebreak CSF2} 
& \multirow{4}{1.6cm}{\centering $1.9 \times 10^{-17}$\linebreak $3.2 \times 10^{-18}$\linebreak $3.0 \times 10^{-16}$\linebreak $3.0 \times 10^{-16}$} 
& \multirow{4}{1.6cm}{\centering 49$\%$\linebreak 33$\%$\linebreak 98$\%$\linebreak 84$\%$ -- 87$\%$}	
& \multirow{4}{2cm}{\centering \cite{Lisdat2016}\linebreak \cite{Huntemann2016}\linebreak \cite{wey18, comm1} \linebreak \cite{wey18, comm1}} \\
\rule[-2ex]{0pt}{8ex}\\

\br
\end{tabular}
\end{center}
\end{table*}

An overview of all the primary and secondary frequency standards involved is given in table~\ref{tab:clock_prop}, with typical values of their systematic uncertainties and with their uptimes during the campaign. The uptime coverage of the optical clocks refers to the whole 26-day interval, and duty cycles up to $77\,\%$ were achieved. For the combination of two optical clocks at one institute, i.e. when at least one clock out of two was operated, duty cycles of $88\,\%$ were reached. The uptime coverage of the fountain clocks refers to intervals of 12 and 16 days in length that were used for the analysis (see sect.~\ref{sec:ana}).

The short- and long-term frequency instability characteristics of the hydrogen masers can be found in table~\ref{tab:HM_prop}. At LNE-SYRTE, the HM is filtered by a microwave oscillator to improve the short term stability of the local clock measurements against the HM.

\begin{table}
\caption{\label{tab:HM_prop} Overview of the hydrogen masers used in the campaign and their typical instabilities at short (1\,s) and long (1\,d) averaging times. The short-term instability of HM-889 at LNE-SYRTE is superior due to filtering with a microwave oscillator.} 
\begin{center}
\footnotesize
\lineup

\begin{tabular}{@{}cccc}
\br
Institute & HM & $\sigma_y(\tau=1\,\mathrm{s})$	& $\sigma_y(\tau=1\,\mathrm{d})$ \\ \ms 
\mr

\multirow{1}{1.8cm}{\centering INRIM} 
& \multirow{1}{1.5cm}{\centering HM3} 
& \multirow{1}{1.5cm}{\centering $3\times 10^{-13}$} 
& \multirow{1}{1.5cm}{\centering $4\times 10^{-16}$}	\\
\rule[-1ex]{0pt}{1ex}\\

\multirow{1}{1.8cm}{\centering LNE-SYRTE} 
& \multirow{1}{1.5cm}{\centering HM-889} 
& \multirow{1}{1.5cm}{\centering $3 \times 10^{-15}$} 
& \multirow{1}{1.5cm}{\centering $5 \times 10^{-16}$}	\\
\rule[-1ex]{0pt}{1ex}\\

\multirow{1}{1.8cm}{\centering NPL} 
& \multirow{1}{1.5cm}{\centering HM2} 
& \multirow{1}{1.5cm}{\centering $5 \times 10^{-13}$} 
& \multirow{1}{1.5cm}{\centering $7 \times 10^{-16}$}	\\
\rule[-1ex]{0pt}{1ex}\\

\multirow{1}{1.8cm}{\centering PTB} 
& \multirow{1}{1.5cm}{\centering H9} 
& \multirow{1}{1.5cm}{\centering \centering $1 \times 10^{-13}$} 
& \multirow{1}{1.5cm}{\centering $5 \times 10^{-16}$}\\

\br
\end{tabular}
\end{center}
\end{table}


The clock comparisons must account for the relativistic redshifts of the clock frequencies, which depend on the gravity (gravitational plus centrifugal) potential experienced by each clock. In this context, the spatial variations of the gravity potential are most important, while temporal variation effects (mainly due to solid Earth and ocean tides) for the clocks are below a few parts in $10^{17}$~\cite{voi16} and have been neglected in this work. Two classical geodetic methods exist to determine the static (spatially varying) potential. These are the geometric levelling approach (levelling together with gravity measurements along the levelling path) and the GNSS/geoid approach, using GNSS positions (ellipsoidal heights) and a high-resolution gravimetric (quasi)geoid model. The GNSS/geoid approach was chosen because it is not affected by systematic levelling errors over long distances. An improved European gravimetric (quasi)geoid model (EGG2015), incorporating new gravity measurements around all clock sites, was utilized in the first instance for the computation of the geopotential numbers for specific existing reference markers near the clock sites ($C_\mathrm{RefMk}$). In addition to this, local levelling measurements were carried out to transfer the geopotential numbers from the reference markers to the actual clock locations, giving $C_\mathrm{clock} = C_\mathrm{RefMk} + g \Delta H$, where $g$ is the local gravity acceleration, and $\Delta H$ is the height difference between the clock point and the reference marker (with $\Delta H = H_\mathrm{clock} - H_\mathrm{RefMk}$). Finally, the fractional relativistic redshift is computed according to
\begin{equation}
\frac{\Delta \nu}{\nu} = \frac{C_\mathrm{clock}}{c^2},
\end{equation}
where $c$ is the speed of light in vacuum. All relevant values are provided in table~\ref{tab:GRS}, where the underlying coordinate reference frame is ITRF2008 with the epoch 2005.0, and the zero reference potential (consistent with the IAU recommendations from the year 2000 and with the resolution 2 of the 26$^\mathrm{th}$ CGPM~2018) for the geopotential numbers is $W_0 = 62\,636\,856.00 \,\mathrm{m}^2\mathrm{s}^{-2}$.
The uncertainties of the differential redshift determinations for the distant clocks are less than $4 \times 10^{-18}$, corresponding to less than 4~cm uncertainty in height~\cite{Denker2018, meh18, del19}.

\begin{table*}
\caption{\label{tab:GRS} 
Data used to derive the relativistic redshifts of the clock frequencies. $C_\mathrm{RefMk}$ is the geopotential number for a specific nearby reference marker based on the geodetic GNSS/geoid approach ~\cite{Denker2018, del19}, $\Delta H$ is the local height difference between the clock and a specific reference marker with $\Delta H = H_\mathrm{clock} - H_\mathrm{RefMk}$, $g$ is the local acceleration due to gravity derived from an FG5-X absolute gravimeter measurement (with an uncertainty far below the significant digits quoted here), and $C_\mathrm{clock}$ is the final geopotential number for the clock point. The underlying coordinate reference frame is ITRF2008 with the epoch 2005.0, and the zero reference potential for the geopotential numbers is $W_0 = 62\,636\,856.00 \,\mathrm{m}^2\mathrm{s}^{-2}$. The uncertainties for $\Delta H$ and for the geopotential numbers $C_\mathrm{clock}$ are given in parentheses, the uncertainty of $C_\mathrm{RefMk}$ is 0.22~m$^2$s$^{-2}$. 
} 
\begin{center}
\footnotesize
\lineup

\begin{tabular}{@{}cccccccc}
\br

\multirow{3}{1.2cm}{\centering Institute} 
& \multirow{3}{1.5cm}{\centering Clock} 
& \multirow{3}{1.3cm}{\centering Reference marker ID} 
& \multirow{3}{2.8cm}{\centering Geopotential number for reference marker $C_\mathrm{RefMk}$	[m$^2$s$^{-2}$]} 
& \multirow{3}{1.3cm}{\centering Height difference $\Delta H$ [m]} 
& \multirow{3}{1.8cm}{\centering Local gravity acceleration \linebreak $g$ [ms$^{-2}$]}	
& \multirow{3}{2.2cm}{\centering Geopotential number for clock $C_\mathrm{clock}$ [m$^2$s$^{-2}$]}
& \multirow{3}{1.8cm}{\centering Redshift correction \linebreak $\frac{\Delta \nu}{\nu} [10^{-15}]$} \\
\rule[-1ex]{0pt}{5ex}\\

\ms 
\mr

\multirow{1}{1.2cm}{INRIM} 
& \multirow{1}{1.5cm}{\centering ITCsF2} 
& \multirow{1}{1.3cm}{\centering CS104} 
& \multirow{1}{2.8cm}{\centering 2323.32} 
& \multirow{1}{1.3cm}{\centering 1.020(10)} 
& \multirow{1}{1.8cm}{\centering 9.8053}	
& \multirow{1}{2.2cm}{\centering 2333.32(24)}
& \multirow{1}{1.8cm}{\centering -25.9617(27)} \\
\rule[-1ex]{0pt}{1ex}\\

\multirow{4}{1.2cm}{\centering LNE- \linebreak SYRTE} 
& \multirow{4}{1.5cm}{\centering Sr2 \linebreak FO1 \linebreak FO2-Cs \linebreak FO2-Rb} 
& \multirow{4}{1.3cm}{\centering SR2 \linebreak FO1 \linebreak FO2 \linebreak FO2} 
& \multirow{4}{2.8cm}{\centering 545.06 \linebreak 613.42 \linebreak	579.63 \linebreak	579.63} 
& \multirow{4}{1.3cm}{\centering 0.201(5)\linebreak 0.761(10) \linebreak 0.962(10) \linebreak 0.886(10)}
& \multirow{4}{1.8cm}{\centering 9.8093}	
& \multirow{4}{2.2cm}{\centering 547.03(23) \linebreak 620.89(24) \linebreak 589.06(24) \linebreak 588.32(24)}
& \multirow{4}{1.8cm}{\centering -6.0865(25) \linebreak -6.9083(27) \linebreak -6.5542(27) \linebreak -6.5459(27)} \\
\rule[-1ex]{0pt}{12ex}\\

\multirow{2}{1.2cm}{NPL} 
& \multirow{2}{1.5cm}{\centering $^{87}$Sr \linebreak $^{171}$Yb$^+$ E3} 
& \multirow{2}{1.3cm}{\centering G4L10 \linebreak G4L16} 
& \multirow{2}{2.8cm}{\centering 96.54 \linebreak 96.58} 
& \multirow{2}{1.3cm}{\centering 1.290(10) \linebreak 1.081(5)}
& \multirow{2}{1.8cm}{\centering 9.8118}	
& \multirow{2}{2.2cm}{\centering 109.20(24) \linebreak 107.19(23)}
& \multirow{2}{1.8cm}{\centering -1.2150(27) \linebreak -1.1926(25)} \\
\rule[-1ex]{0pt}{6ex}\\

\multirow{4}{1.2cm}{PTB} 
& \multirow{4}{1.5cm}{\centering $^{87}$Sr \linebreak $^{171}$Yb$^+$ E3 \linebreak CSF1 \linebreak CSF2} 
& \multirow{4}{1.3cm}{\centering PB02 \linebreak KB02 \linebreak KB02 \linebreak KB02} 
& \multirow{4}{2.8cm}{\centering 763.84 \linebreak 753.04 \linebreak 753.04 \linebreak 753.04} 
& \multirow{4}{1.3cm}{\centering 0.538(3) \linebreak 1.001(3) \linebreak 1.632(5) \linebreak 1.526(5)}
& \multirow{4}{1.8cm}{\centering 9.8125}	
& \multirow{4}{2.2cm}{\centering 769.12(22) \linebreak 762.86(22) \linebreak 769.05(23) \linebreak 768.01(23)}
& \multirow{4}{1.8cm}{\centering -8.5576(25) \linebreak -8.4880(25) \linebreak -8.5569(25) \linebreak -8.5453(25)} \\
\rule[-1ex]{0pt}{8ex}\\

\br
\end{tabular}
\end{center}
\end{table*}

\subsection{TWSTFT Setup}
For the broadband TWSTFT experiment, the ground station equipment at each site comprised a very small aperture terminal (VSAT) roof station and a satellite time and ranging equipment modem (SATRE modem, TimeTech GmbH).
The HM signal was directly used as the reference for the modem.
As the SATRE modems are usually equipped with a maximum of two receive (RX) channels, two modems were daisy-chained to allow all three remote stations to be tracked whilst simultaneously performing a satellite ranging measurement.
The modems were operating at the maximum modulation chip rate of 20\,Mchip/s, requiring a total bandwidth of 36\,MHz, i.e. one full Ku-band satellite transponder.
For this experiment, transponder capacity on satellite ASTRA 3B at 23.5$^{\circ}$E, operated by SES (Soci\'et\'e Europ\'eenne des Satellites), was leased. 

The principle of a TWSTFT measurement ~\cite{ITU_TW}, i.e. the bidirectional exchange of pseudo-random code signals modulated onto a microwave carrier between two remote ground stations over a satellite, results in a compensation of the overall transit times originating from the signal paths between the stations and the satellite.
However, when performing TWSTFT measurements at high resolution, various influences that cause small violations of the reciprocity of the signal paths will become significant and have to be taken into account.
Since frequencies were compared and no absolute phase measurement was carried out, only the temporal variations of these effects need to be considered here.

In two-way frequency transfer, geometric terms need to be taken into account, i.e. the variation of the Sagnac effect and the variation of the path delay difference, while the gravitational terms cancel~\cite{Petit1994}.  Both geometric terms arise from the residual motion of the satellite around its nominal geostationary position, a daily oscillation with a maximum amplitude of about 30\,km. In order to compensate for the variations of the path delay difference, delays with respect to the local UTC(\textit{k}) time scales were introduced to the reference 1 pulse per second (PPS) signals to ensure all signals arrive at about the same time at the satellite. The variations of the Sagnac effect and the residual variations of the path delay difference were then calculated using satellite position and velocity data provided by SES. This results in peak-to-peak relative frequency corrections between $8 \times 10^{-16}$  (LNE-SYRTE -- INRIM) and $2 \times 10^{-15}$ (PTB -- NPL), with the Sagnac term being dominant over the residual path delay difference contribution.

Another influence on the signal path delay is the dispersion of the atmosphere.
The tropospheric contribution can be neglected because of its low frequency dependence and the short signal path, due to the high elevation of the satellite~\cite{ITU_TW, Piester2007}.
For the ionosphere, the frequency dependence of the signal transit time is larger, and a correction can be calculated based on the total electron content (TEC) of the ionosphere~\cite{ITU_TW}.
For the calculation, the values provided in~\cite{TEC_data} were used, and the projection mapping function as described in~\cite{Xubook2003} was applied to estimate the extension of the signal path due to the slant propagation path through the ionosphere.
The peak-to-peak relative frequency corrections due to variation of the difference of ionospheric delays were about $1.3\times10^{-15}$ for the link between PTB and INRIM (maximum influence of the ionosphere) and $4\times10^{-16}$ for the link LNE-SYRTE -- NPL (minimum influence of the ionosphere).

\subsection{GPS Setup}\label{subsec:exp_gps}
As for the TWSTFT ground stations, the GPS ground station equipment regularly employed for contributions to TAI was used, consisting of a GNSS receiver and a GNSS choke ring antenna connected by a coaxial cable with a length between 30\,m and 50\,m.
The receivers were Javad Legacy at INRIM, Septentrio PolaRx4-TR at LNE-SYRTE, DICOM GTR50 at NPL, and Septentrio PolaRx4-TR at PTB. 
Like the TWSTFT modems, the GPS receivers were directly referenced to the hydrogen masers.
All receivers provided data in the RINEX format (versions 2.10 and 2.11), enabling PPP processing. The PPP algorithm by the National Resources Canada (NRCan)~\cite{kou2001} was used for the computation of GPS link phase data. It takes into account various effects such as the propagation through the troposphere (based on models and meteorological data) and ionosphere, antenna phase centre variations~\cite{sch2007}, carrier-phase windup~\cite{wu1993}, relativistic effects~\cite{ash2003}, and site displacements, e.g. due to solid Earth tides. 

\section{Clock comparison data analysis}\label{sec:ana}

As usual in time and frequency metrology, we denote the fractional frequency offset from the nominal frequency as ``relative frequency deviation'' (symbol $y$) and use the symbol $x$ for the time deviation associated with the phase~\cite{sul90}.  
The calculation of the relative frequency deviation difference between two remote atomic clocks located at laboratories 1 and 2, respectively, involves two frequency data sets (hydrogen maser (HM) against atomic clock (AC) at each institute, called ``clock data'' in this paper) and one phase data set for the link (remote comparison of hydrogen masers, ``link data'' in this paper):

\begin{eqnarray}
& \quad y^\mathrm{c1}(t) = y^\mathrm{AC1}(t) - y^\mathrm{HM1}(t) \label{eq:y1_dataset} \\
& \quad y^\mathrm{c2}(t) = y^\mathrm{AC2}(t) - y^\mathrm{HM2}(t) \label{eq:y2_dataset} \\
& \quad x^\mathrm{link}(t) = x^\mathrm{HM1}(t) - x^\mathrm{HM2}(t). \label{eq:x_dataset}
\end{eqnarray}

\noindent{If the relative frequency deviation data sets $y(t)$ are represented by a uniform time series $y_i=y(t_i)$ with sampling times $t_i$ on an equispaced time grid with steps $\Delta t$, and the phase data set $x^\mathrm{link}(t)$ is represented by a corresponding time series $x^\mathrm{link}_i=x^\mathrm{link}(t_i)$, it can be converted into a relative frequency deviation time series:}

\begin{eqnarray}
& y^\mathrm{link}_i & = \frac{x^\mathrm{link}_{i+1} - x^\mathrm{link}_i}{\Delta t} \nonumber \\
& & = y^\mathrm{HM1}_i - y^\mathrm{HM2}_i.
\label{eq:x_y_convert}
\end{eqnarray}

\noindent{The time series of relative frequency deviation differences between the remote clocks can be calculated as:}

\begin{eqnarray}
 y^\mathrm{AC1-AC2}_i & = y^\mathrm{AC1}_i - y^\mathrm{AC2}_i \nonumber \\
 & = y^\mathrm{c1}_i - y^\mathrm{c2}_i + y^\mathrm{link}_i,
\label{eq:y_merge}
\end{eqnarray} 

\noindent{from which we could directly compute the mean value and its statistical uncertainty.} 

\begin{figure*}[]
\centering
\includegraphics[width=2\columnwidth]{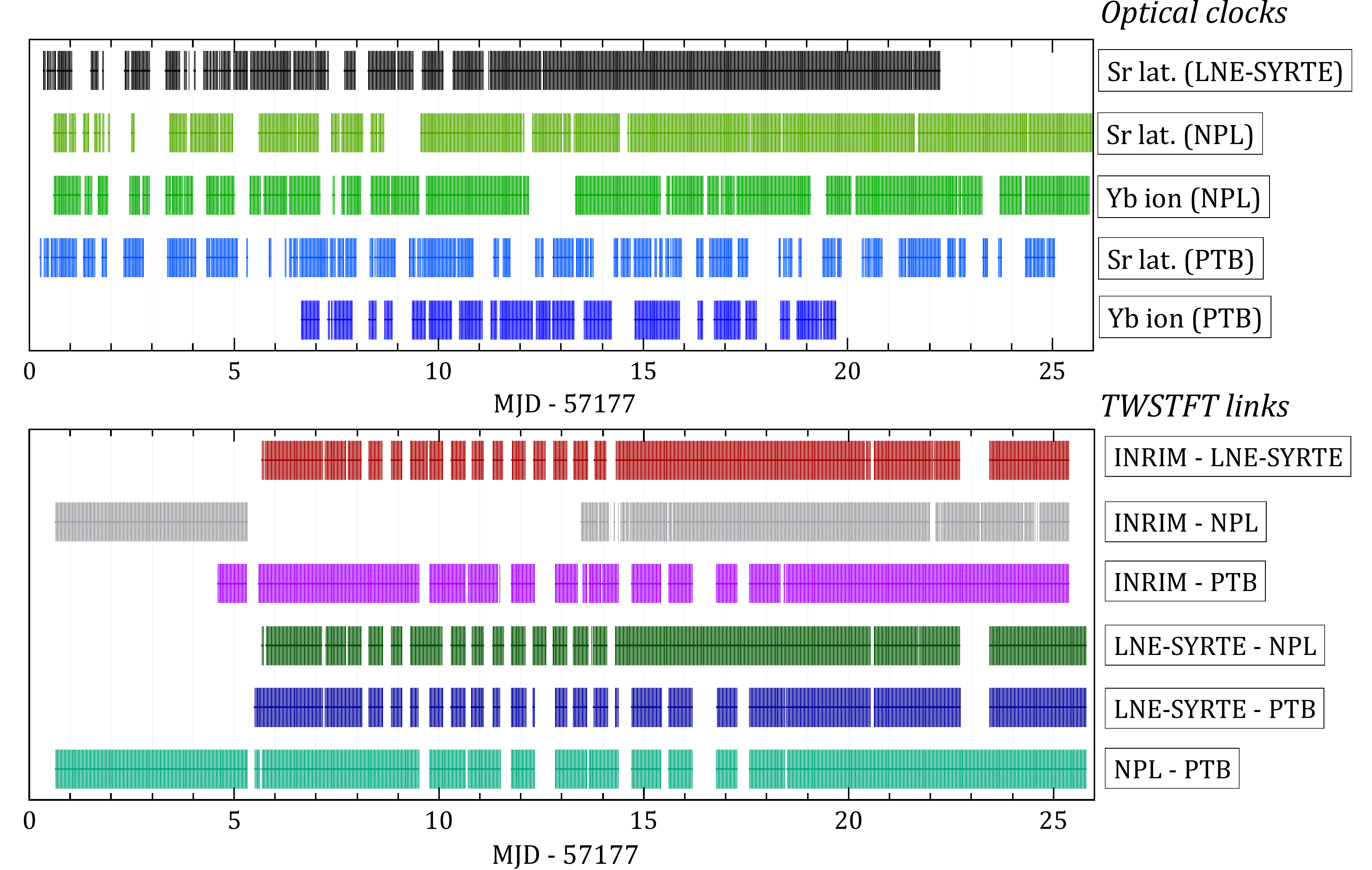}
	\caption{Measurement times during the campaign of all optical clocks involved (above) and the TWSTFT links (below). Periods with technical disturbances resulting in outliers have been excluded. The GPS links were continuously operated, but the intervals used for the clock comparisons were limited by discontinuities arising from technical disturbances. Hence, data from the GPS receivers at INRIM and LNE-SYRTE were only available for the intervals 57177.0 -- 57198.0 and 57183.0 -- 57197.0. The numbers refer to to Modified Julian Date (MJD), which is based on a continuous count of days, in which 57177.0 corresponds to 4th June 2015, 00:00:00 UTC.} 
	\label{fig:uptime_clocks}
\end{figure*}

\begin{figure*}[]
\centering
\includegraphics[width=2\columnwidth]{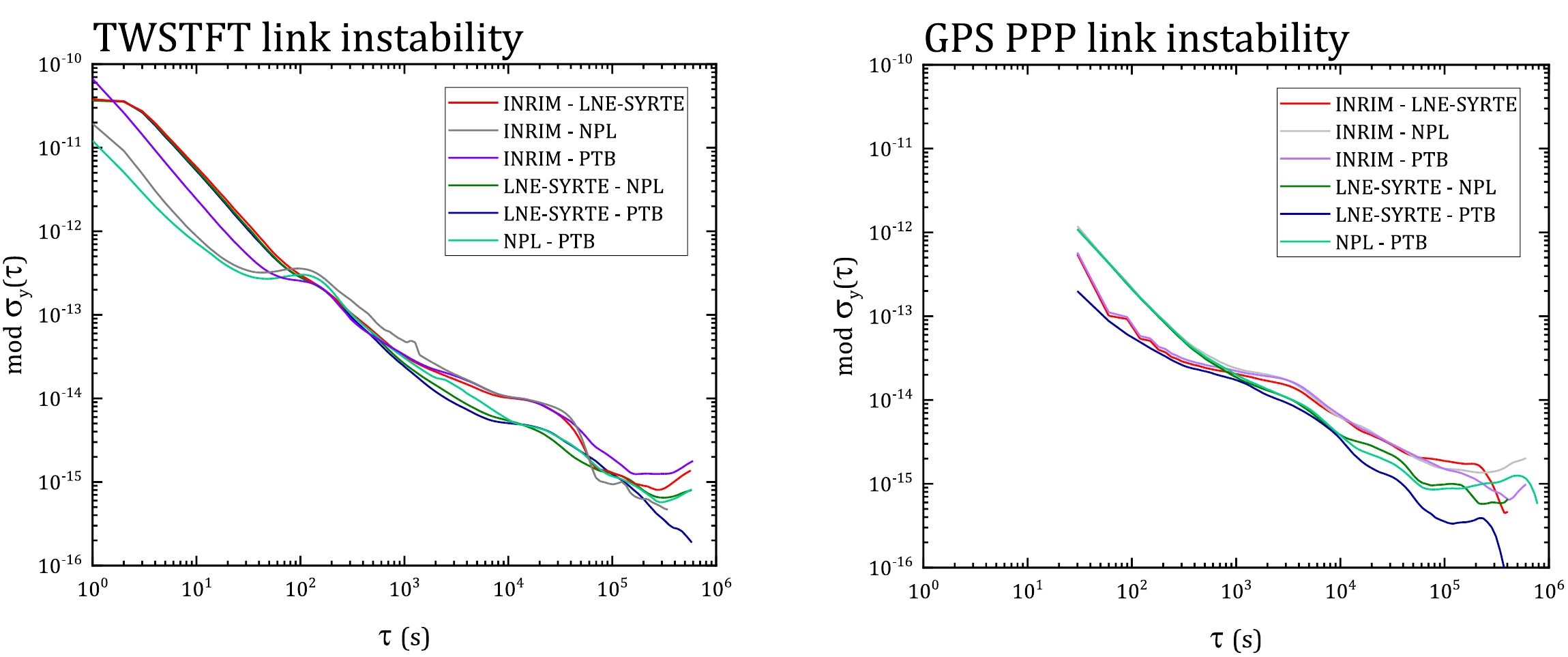}
	\caption{Modified Allan deviation of all link data, calculated over the whole measurement period. For averaging times below 1\,day, the modified Allan deviation reflects the frequency instability of the links. Above 1\,day, the modified Allan deviations are dominated by the HM fluctuations.}
	\label{fig:Mdev_all_links}
\end{figure*}

However, the measurement data in practice have different start and end times, and have gaps at different times and of different durations (as seen in fig.~\ref{fig:uptime_clocks} for the gaps in the optical clock and the TWSTFT data during the campaign). They are thus not straightforwardly representable by uniform time series as assumed above. In addition they have different noise properties: the clock data are dominated by the noise of the HMs, whereas the link data are dominated by the satellite link noise up to averaging times of about 0.5\,d, as can be seen in the modified Allan deviations of the TWSTFT link data (see fig.~\ref{fig:Mdev_all_links}). Only at longer averaging times does the HM noise prevail together with other coloured noise contributions on the link, most probably caused by environmental influences such as temperature variations.

Different boundary conditions are given in the cases of optical and caesium fountain clocks. For the optical clocks, the data are available on a 1\,s grid, and they contain a relatively large number of gaps. The  uncertainties of the optical clocks are known from local comparisons to range below $10^{-16}$, so in this case, it is necessary to suppress all disturbances which can deteriorate the remote clock comparison uncertainty. 

The fountain clock comparisons were operated without any major interruptions, but the concept of operation provides frequency values at intervals which do not in general correspond to integer seconds. Furthermore, the statistical uncertainty of the fountain clocks for averaging times of 1\,d is in the few 10$^{-16}$ range.

For these reasons we applied different analysis procedures to the optical clock comparisons and to the fountain clock comparisons, although both methods utilize some kind of pre-averaging of the link phase data in order to suppress the link noise.

\subsection{Determination of the relative frequency deviation difference between pairs of optical clocks (OC)}
\label{sec:OC_data_analysis}
If we were to naively apply eqs.~\ref{eq:y1_dataset} -- \ref{eq:y_merge}, the instability and the resulting statistical uncertainty would be limited by the unsuppressed white phase noise of the link and possibly by a residual contribution from HM fluctuations in the presence of gaps in the phase and frequency datasets, to a level much larger than $10^{-16}$. In fact, due to the white phase noise, the instability would not be limited to approximately $5\times 10^{-10}(\tau/\mathrm{s})^{-1}$ as in fig.~\ref{fig:Mdev_all_links}, but rather to $5\times 10^{-10}(\tau/\mathrm{s})^{-1/2}$, which would render the remote optical clock comparisons useless. Hence, the link data should be pre-averaged to suppress the link noise, and the different data sets should be combined in a way that suppresses the HM fluctuations. Such an analysis procedure is described in the following.

According to the \textit{Guide to the expression of Uncertainty in Measurement} (GUM)~\cite{gum95}, the standard deviation $\sigma _{\bar{y}}$ of the mean describes the type-A uncertainty of the mean. 
The choice of the estimator for $\sigma _{\bar{y}}$ depends on knowledge of the noise types involved, or in other words of serial correlations in the relative frequency deviation data. The commonly used estimator

\begin{equation}
\hat \sigma_{\bar{y}} = \sqrt{\frac{\sum_{i=1}^{N} (y_i - \bar{y})^2}{N(N-1)}}
\label{eq:s_mean_bias}
\end{equation}
\noindent{ignores correlations and is therefore only unbiased in the case of white frequency noise. If other noise contributions are present it can significantly underestimate the standard deviation of the mean.}
In the optical clock comparisons via satellite link we have a mixture of coloured noise processes, i.e. rather intricate correlations. Furthermore, the relative frequency deviation data contain gaps. Therefore, we derive an alternative approach for an estimator of $\sigma _{\bar{y}}$ (see  appendix), which is based on ideas from~\cite{Zhang2006, Zieba2010, Zieba2011}. This takes into account both the correlations and the data gaps. The correlations are included via autocovariances, which can themselves be estimated from the relative frequency deviation data. The gaps are handled by allowing for weighting of the relative frequency deviation data and setting the weighting factors to zero at gaps. For this purpose, the relative frequency deviation data are represented by a time series $y_i$ on a uniform time grid and with weighting factors $w_i$, with the index $i=1\dots N$, where $N$ is the total number of elements of the time series. As derived in the appendix, an estimator for the standard deviation of the weighted mean is:
\begin{equation}
\hat \sigma_{\bar{y}}^2 = {\sum_{i=1}^{N} w_i (y_i - \bar{y})^2} + 2 \sum_{l=1}^{l_\mathrm{cut}} \hat{R}_{l} \sum_{j=1}^{N-l} \sqrt{w_j w_{l+j}},
\label{eq:s_mean_unbias}
\end{equation}
\noindent{where $\hat{R}_{l}$ is the estimator of the lag-$l$ autocovariance given by equation~\ref{eq:Rhat_l} in the appendix, and $l_\mathrm{cut}$ is a cutoff index.} The latter is introduced because the autocovariance can only be calculated with sufficient accuracy for lower lags. At higher lags the coefficients become erroneous~\cite{Zieba2011}. 

\begin{figure*}[]
\centering
\includegraphics[width=170mm]{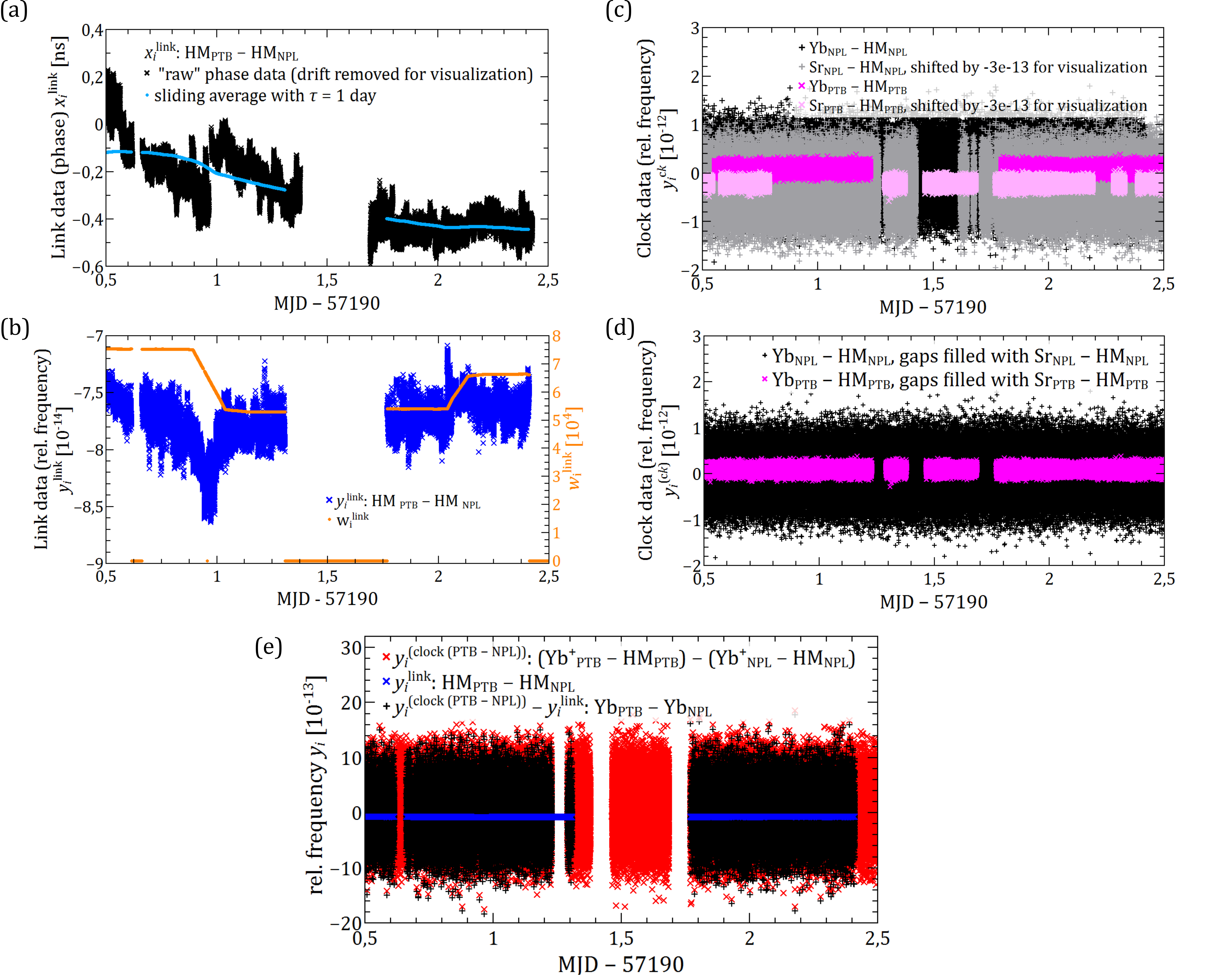}
\caption{Scheme of the data analysis procedure, shown for the comparison of Yb$^+$(PTB) and Yb$^+$(NPL) via TWSTFT for two days of the campaign. (a) Phase data of the link between the hydrogen masers at PTB and NPL, with the phase drift (frequency offset) over the whole campaign interval removed for better visualization. Black symbols: Measured data (outliers removed), blue dots: Moving average with a window of 1\,day. The time points of the pre-averaged phase data are different from those of the measured phase data (centres of gravity as described in the text). For this reason, the gap edges are different for the black and blue curve. (b) Blue symbols: Relative frequency deviation calculated from the averaged phase data. Orange: Weights $w_i^\mathrm{link}$. (c) Locally measured differences between the hydrogen maser and the optical clocks (pink: PTB, black/grey: NPL). Dark symbols: Yb$^+$ clocks, light symbols: Sr clocks. The NPL data show increased noise in comparison to the PTB data because of the different short-term stability of the hydrogen masers (see table~\ref{tab:HM_prop}). The Sr clock data are used to fill the gaps in the Yb$^+$ data at the same institute, which results in the graphs shown in (d). (e) The final step is depicted. The two time series of (d) are merged into one $y^\mathrm{clock(PTB-NPL)}_i$ (red), and in blue the frequency difference of the hydrogen masers (blue graph from (b)) is shown. Both time series are subtracted in order to get the frequency difference of the two Yb$^+$ clocks (black symbols).}
	\label{fig:sample_data}
\end{figure*}

With this tool we can now develop an analysis strategy for the optical clock comparisons. The determination of the relative frequency deviation difference from the three time series $x^\mathrm{link}_i$, $y^\mathrm{OC1}_i$ and $y^\mathrm{OC2}_i$ for the example optical clock comparison $^{171}\mathrm{Yb}^+$(NPL) versus $^{171}\mathrm{Yb}^+$(PTB) is illustrated by figure~\ref{fig:sample_data}.

There are two opposing requirements which should be satisfied, but which can both degrade the statistical uncertainty. Hence, we have to find a compromise which minimizes the statistical uncertainty.
On the one hand, we should combine clock- and link-data only at times at which they overlap, in order to achieve the best common mode rejection of the HM fluctuations.
On the other hand, the phase data should be pre-averaged, and with regard to the uncertainty determination, a time series with a small time grid is desirable in order to estimate the autocovariance reliably. 

We apply a sliding arithmetic mean to the phase data and convert the pre-averaged phase data into frequency data by using eq.~\ref{eq:x_y_convert} with $\Delta t$~= 1\,s. As a negative consequence, HM fluctuations during gaps in the data are partially folded to adjacent points of time, at which the link frequency data is subsequently combined with the clock data. Hence the common mode rejection of the HM fluctuations is not perfect any more. For the sliding average of the link phase data, an averaging window length of 1 day is chosen as a compromise between the two opposing aspects: diurnal oscillations as well as other disturbances present on the link phase data at averaging times between 0.5 and 1\,d are suppressed along with the white and flicker phase noise, without introducing systematic errors, while the HM fluctuations at up to 1\,d are still tolerable.

Three details should be noted in connection with the conversion of the pre-averaged phase data to frequency data, especially when there are gaps in the phase data:
\newline
First, we use a time step of $\Delta t$~= 1\,s for the conversion with eq.~\ref{eq:x_y_convert}. Larger intervals would introduce artificial correlations at larger lags $l > l_\mathrm{cut}$, which could ultimately yield a biased statistical uncertainty. At gaps, however, the introduction of such correlations cannot be totally avoided. In order to minimize their influence, we cut the original gaps of the raw link phase data into the pre-averaged data, which also reduces the negative influence on the common mode rejection of the HM fluctuations discussed above.
\newline
Second, if the data are asymmetrically distributed within the phase data averaging window due to gaps, the effective point of time at which the phase average is valid is not in the centre of the averaging window, but instead at the temporal centre of gravity. For this reason, the temporal centres of gravity rounded to values on the time series grid are used instead of the original times for the conversion to frequency values.
\newline
Third, the number $n_i$ of phase values $x_i$ falling within the averaging window varies near gaps, and larger weights should be assigned to frequency values derived from a larger number of phase values. Hence, we derive semi-heuristic weights for the link frequency data from the time series of the $n_i$:
\begin{equation}
w_i^\mathrm{link}= \frac{1}{\frac{1}{n_i}+\frac{1}{n_{i+1}}}=\frac{n_i n_{i+1}}{n_i + n_{i+1}}.
\label{eq:linkweights}
\end{equation}

\noindent{These weights are not yet normalized to unity, but can be interpreted as an effective number of frequency values in the 1\,day phase averaging window. This is motivated by the fact that each frequency value $y^\mathrm{link}_i$ on the  1\,s grid is determined from the difference between the 1\,day averages $\overline{x}_i$ and $\overline{x}_{i+1}$ over $n_i$ and $n_{i+1}$ phase values, respectively, divided by the difference between the time centres of gravity.}
In fig.~\ref{fig:sample_data}\,(a), the original phase data and the averaged phase data are shown for a two-day interval. The resulting relative frequency deviation and the associated weights $w_i^\mathrm{link}$ are visualized in fig.~\ref{fig:sample_data}\,(b).

We should also avoid the introduction of artificial correlations on the clock data, so we do not pre-average $y^\mathrm{c1}$ and $y^\mathrm{c2}$. However, when taking only the data for which $y^\mathrm{link}$, $y^\mathrm{c1}$ and $y^\mathrm{c2}$ overlap, we would discard a lot of information. Hence, if an institute operates two clocks (say A and B), we fill the gaps in the data of clock A, whenever clock B is running while clock A is not and vice versa. For this purpose, we use the ratio between clock A and B measured with small uncertainty during the periods when A and B are operative during the campaign (~\ref{fig:sample_data}\,(c) and (d)).
This is justified, because the uncertainty of the local ratio is negligible with respect to the uncertainty of the remote comparison. This approach has been applied before in an optical clock comparison between NICT in Japan and PTB in Germany~\cite{Hachisu2014}.
Just as for the link data, weights, not yet normalised, are assigned to each clock data frequency value:
\begin{equation}
w_i^{y^\mathrm{c}}=\left\{ \begin{array}{ll} 
				0 		&  \mathrm{at\, gaps} \\
				1  				& \mathrm{ elsewhere}.
				\end{array} \right.
\label{eq:clockweights}
\end{equation}

\noindent{In the next step, the three resulting frequency data time series are merged according to eq.~\ref{eq:y_merge}. An overall weight was assigned to each frequency value, given by the product of the individual weighting factors and normalized to unity:}
\begin{equation}
w_i=\frac{w_i^{y_1} w_i^{y_2} w_i^\mathrm{link}}{\sum_{i=1}^N w_i^{y_1} w_i^{y_2} w_i^\mathrm{link}}.
\label{eq:weights}
\end{equation}

\noindent{The time series $y^\mathrm{link}_i$, ($y^\mathrm{c1}_i - y^\mathrm{c2}_i$) and $y^\mathrm{OC1-OC2}_i$ are depicted in fig.~\ref{fig:sample_data}\,(e).}
From this, a weighted mean of the relative frequency deviation difference between the two optical clocks located at laboratories 1 and 2 $\bar y^\mathrm{OC1-OC2}$ was calculated for the overall campaign. 

In the case of GPS PPP data, the time series of the link data is on a $\Delta t$~= 30\,s grid.
The procedure was adapted accordingly.

\subsubsection*{Determination of uncertainties.} 

The statistical uncertainty of the relative frequency deviation difference between two optical clocks is calculated as described above with eq.~\ref{eq:s_mean_unbias}. 
We chose a cutoff index $l_\mathrm{cut} = 4000$. Above this lag, no significantly non-zero values of the autocovariance estimator were observed.

As known from local optical clock comparisons, the contribution of the optical clocks to the statistical uncertainty is well below $10^{-16}$ and thus negligible in the remote comparison.

For the systematic uncertainty, the considered contributions originate from both the clocks and the link. To determine the systematic uncertainty of the satellite links, various influences were investigated, with the temperature variations both inside the laboratory and outdoor near the antennas being the dominant contributions.
During the campaign the relevant temperatures were monitored at all stations. To estimate possible frequency errors due to the observed temperature changes, phase deviations $x_\mathrm{T,i}$ were deduced from the recorded temperature data and temperature sensitivities $c_\mathrm{T,i}$ for individual components $i$ of the link equipment. With these phase data, the same analysis process was performed as for the raw link phase data, resulting in the component-related temperature induced frequency shift $\bar{y}_\mathrm{T,i}$. Because of the uncertainty in the temperatures and sensitivities of the various components, we quadratically add all individually inferred frequency shifts and use the result as estimate for the systematic uncertainty of the link. 
For the TWSTFT links, measurements of the temperature coefficients for both indoor and outdoor equipment were performed before the campaign. Since not all components of every base station could be tested, the largest temperature sensitivities observed for the class of components were used for a conservative estimation: $c_\mathrm{T}$~= 5\,ps/K for the SATRE modems and $c_\mathrm{T}$~= 5\,ps/K for the combination of frequency converters and amplifier.

For the GPS measurement analysis, temperature coefficients for the main components (receiver, antenna and antenna cable) were taken from the literature for the analysis~\cite{Prillaman2010, Ray2003, Weinbach2013a}: 9\,ps/K for the PolaRX4 receivers, 13\,ps/K for the GTR50, 0.03\,ps/K$\cdot$m for the antenna cables and 10\,ps/K for the antennas. 

Details about the systematic uncertainties of the individual optical and fountain clocks can be found in the references in table~\ref{tab:clock_prop}.

\subsection{Determination of the relative frequency deviation difference between pairs of fountain clocks}

As mentioned above, the situation is different for the fountain clock comparisons. 
The high duty cycles of the fountain clocks allow for the approach of determining average frequency values for each time series separately (eq.~\ref{eq:y1_dataset} to ~\ref{eq:x_dataset}). 
So, we first identified two appropriate measurement intervals common for fountain clocks and links of at least two of the partners INRIM, LNE-SYRTE and PTB. These intervals chosen for analysis are:

\begin{itemize}
\item MJD 57183.0 -- MJD 57199.0 (16\,d) for broadband TWSTFT for all three links between INRIM, LNE-SYRTE, and PTB;
\item MJD 57184.0 -- MJD 57196.0 (12\,d) for GPS PPP on all three links.
\end{itemize}

Subsequently, the individually measured frequency differences between a fountain and the local hydrogen maser were averaged for such intervals. To evaluate the mean frequency difference between the remote hydrogen masers $\bar{y}^\mathrm{link}$, we calculated the average phase values for the TWSTFT or GPS link over 24 hours at the beginning and at the end of the respective interval, centred around 0:00 UTC of the first/last day of the respective comparison interval. We then obtained the mean frequency difference $\bar{y}^\mathrm{link}$ by dividing the difference between the two phase averages by the duration of the interval ($\Delta t_\mathrm{FC}$, with FC = fountain clock comparison). By this means, the influence of white phase noise and of the other disturbances like diurnals on the link was reduced. For a frequency averaging interval length of 1\,day, this processing is equivalent to $\Lambda$-averaging, but for longer frequency averaging intervals, the frequency weighting function has a trapezoidal shape instead of the $\Lambda$-shape.

The averaged frequency values $\bar{y}^\mathrm{link}$, $\bar{y}^\mathrm{c1}$ and $\bar{y}^\mathrm{c2}$ of the three time series were merged into one relative frequency deviation difference between the remote fountain clocks in analogy to eq.~\ref{eq:y_merge}.

\subsubsection*{Determination of uncertainties}
The determination of the statistical uncertainties of the fountain clocks is based on the measured short-term instability and the known averaging behaviour (white frequency noise) for longer measurement times. The systematic uncertainties have already been presented in section~\ref{sec:clocks}, table~\ref{tab:clock_prop}. In addition, a dead-time uncertainty was taken into account, as generally found in Circular T~\cite{CircT}.

The uncertainty evaluation for the link during the frequency averaging interval required several steps. The fluctuations of the phase data during the first and last day of the frequency averaging interval carry information about the short term $(\le 1\,\mathrm{d})$ link instability. Together with knowledge about the relation between the instability and statistical uncertainty $u_\mathrm{A}$~\cite{Benkler2015}, we can thus derive the uncertainty for a frequency averaging interval length of 1\,day. As mentioned before, for a 1\,day frequency averaging interval the frequencies are $\Lambda$-weighted, so the uncertainty at 1\,day is given by the modified Allan deviation at 1\,day, multiplied by a form factor of $\sqrt{2/3}$ for the link data dominated by white phase noise:
\begin{equation}
u_\mathrm{A}(1\,\mathrm{d}) = \sqrt{2/3}\,\mathrm{mod}\,\sigma_y(\tau=1\,\mathrm{d}).
\label{eq:u_A_modADEV}
\end{equation}
However, at the longer averaging intervals used for the fountain comparisons, we cannot derive such information about the pure link instability solely from the modified Allan deviation estimated from the link data: First, the modified Allan deviation is dominated by the HM fluctuations at $\tau>1\,\mathrm{d}$. Second, the modified Allan deviation can usually be estimated only at averaging times shorter than the total length of the measurement interval. However, we can instead make use of the statistical uncertainties $u_\mathrm{A,\,OC}$ determined in the optical clock comparisons on intervals of length $\Delta t_\mathrm{OC}$ longer than $\Delta t_\mathrm{FC}$, because there the influence of the optical clocks is negligible and $u_\mathrm{A,\,OC}$ is marginally affected by the HMs. Thus, we can interpolate linearly between $u_\mathrm{A}(1\,\mathrm{d})$ derived using eq.~\ref{eq:u_A_modADEV} and $u_\mathrm{A,\,OC}(\Delta t_\mathrm{OC})$, in order to obtain the link uncertainty $u_\mathrm{A,\,FC}(\Delta t_\mathrm{FC})$ at the frequency averaging interval length used for the fountain clock comparison. For the values of $u_\mathrm{A,\,OC}$ and $\Delta t_\mathrm{OC}$, see tables~\ref{tab:results_even} and \ref{tab:results_odd} in the next section.

This procedure was applied directly for the link LNE-SYRTE -- PTB. Without optical clock data at INRIM, however, for the INRIM-related links we used the typical uncertainties $u_\mathrm{A,\,OC}(\Delta t_\mathrm{OC})$ observed with the other links.

The systematic uncertainty of the link was determined in the same way and with the same parameters as for the optical clock comparisons. To complete the information on the temperature coefficients of the GPS equipment, the values used for INRIM were 13\,ps/K for the receiver, 0.5\,ps/K$\cdot$m for the antenna cable and 10\,ps/K for the antenna. 

\section{Results and discussion}\label{sec:res}

\subsection{Comparison of optical clocks}
\begin{figure*}[]
\centering
\includegraphics[width=160mm]{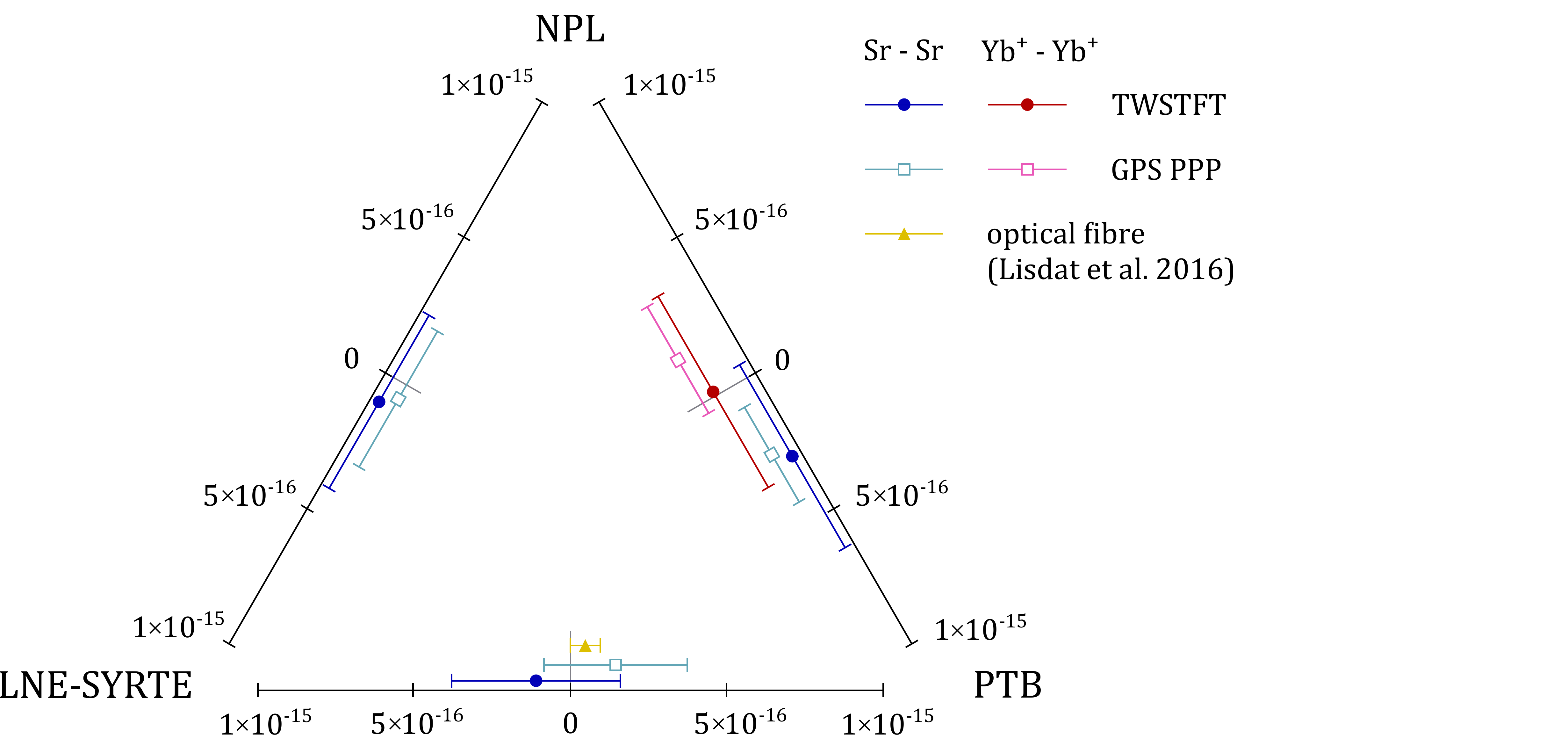}
\caption{The results of the relative frequency deviation differences for the optical clock comparisons between the three institutes LNE-SYRTE, NPL and PTB, for clocks of the same type. The triangle plot is explained in the text at the beginning of section~\ref{sec:res}. The result for the fibre comparison is taken from~\cite{Lisdat2016}.}
\label{fig:results_even}
\end{figure*}

In table~\ref{tab:results_even}, the results for the relative frequency deviation differences between the optical clocks of the same type are listed, with the statistical and systematic uncertainty contributions.
\begin{table*}
\caption{\label{tab:results_even} Results for the comparisons of optical clocks of the same type. The third column gives the average optical frequency difference, $u_\mathrm{B,c}$ is the combined systematic clock uncertainty, $u_\mathrm{A,l}$ and $u_\mathrm{B,l}$ correspond to the statistical and systematic link uncertainties, $u$ is the combined uncertainty, and $\Delta t_\mathrm{OC}$ corresponds to the effective length of the optical clock comparison. All values except for the last column $\Delta t_\mathrm{OC}$ are in $10^{-16}$.} 
\begin{center}
\footnotesize
\lineup

\begin{tabular}{@{}cccccccc}

\br
Clock pair & link & difference & $u_\mathrm{B,c}$	& $u_\mathrm{A,l}$	& $u_\mathrm{B,l}$ & $u$ & $\Delta t_\mathrm{OC}$ [d]\\ \ms 
\mr

\multirow{2}{2.3cm}{\centering Sr(LNE-SYRTE) \linebreak-- Sr(NPL)} 
& \multirow{2}{1.8cm}{\centering TWSTFT\linebreak GPS PPP} 
& \multirow{2}{.6cm}{\centering \- 0.9\linebreak \- 0.5} 
& \multirow{2}{.6cm}{\centering 0.8}	
& \multirow{2}{.6cm}{\centering 3.0\linebreak 2.3}	
& \multirow{2}{.6cm}{\centering 0.7\linebreak 0.8}	
& \multirow{2}{.6cm}{\centering 3.2\linebreak 2.5}
& \multirow{2}{.6cm}{\centering 16.6\linebreak 13.5} \\
\rule[-2ex]{0pt}{8ex}\\

\multirow{2}{2.3cm}{\centering Sr(LNE-SYRTE) \linebreak-- Sr(PTB)} 
& \multirow{2}{1.8cm}{\centering TWSTFT\linebreak GPS PPP} 
& \multirow{2}{.6cm}{\centering \- 1.1\linebreak -1.4} 
& \multirow{2}{.6cm}{\centering 0.5}	
& \multirow{2}{.6cm}{\centering 2.5\linebreak 1.9}	
& \multirow{2}{.6cm}{\centering 0.9\linebreak 1.2}	
& \multirow{2}{.6cm}{\centering 2.7\linebreak 2.3} 
& \multirow{2}{.6cm}{\centering 16.4\linebreak 13.5} \\
\rule[-2ex]{0pt}{8ex}\\

\multirow{2}{2.3cm}{\centering Sr(NPL) \linebreak-- Sr(PTB)} 
& \multirow{2}{1.8cm}{\centering TWSTFT\linebreak GPS PPP} 
& \multirow{2}{.6cm}{\centering -2.9\linebreak -2.5} 
& \multirow{2}{.6cm}{\centering 0.7}	
& \multirow{2}{.6cm}{\centering 3.3\linebreak 1.5}	
& \multirow{2}{.6cm}{\centering 0.5\linebreak 0.6}	
& \multirow{2}{.6cm}{\centering 3.4\linebreak 1.8} 
& \multirow{2}{.6cm}{\centering 24.4\linebreak 24.5} \\
\rule[-2ex]{0pt}{8ex}\\

\multirow{2}{2.3cm}{\centering Yb$^+$(NPL) \linebreak-- Yb$^+$(PTB)} 
& \multirow{2}{1.8cm}{\centering TWSTFT\linebreak GPS PPP} 
& \multirow{2}{.6cm}{\centering \- 0.2\linebreak \- 1.6} 
& \multirow{2}{.6cm}{\centering 1.1}	
& \multirow{2}{.6cm}{\centering 3.3\linebreak 1.5}	
& \multirow{2}{.6cm}{\centering 0.5\linebreak 0.6}	
& \multirow{2}{.6cm}{\centering 3.5\linebreak 2.0}
& \multirow{2}{.6cm}{\centering 24.4\linebreak 24.5} \\
\rule[-1ex]{0pt}{4ex}\\

\br
\end{tabular}
\end{center}
\end{table*}
For a graphical visualization of clock comparison results with clocks located at three institutes, we introduce triangle plots as shown in fig.~\ref{fig:results_even} for clock pairs of the same species. The intention behind such triangle plots is that some important features of a triangle clock comparison might be grasped more intuitively than in the form of a mere table or of a Cartesian scatter plot. For example, it is easier to identify a clock whose frequency deviates from the frequency of the other two clocks. For this purpose, each corner of the triangle corresponds to one institute. Between the corners, there are relative frequency deviation axes, with the relative frequency deviation difference $y=0$ in the centre, which for clock pairs of the same species corresponds to a frequency ratio of 1 between the clocks located at the institutes indicated on the respective corners. If the relative frequency deviation difference differs from 0 (or the frequency ratio deviates from 1), we use the following convention: For $y^\mathrm{(c1-c2)}>0$, the data point is drawn on the axis in the direction of the corner where clock c1 is located. The axes range from zero in the centre to $1\times10^{-15}$ at each of the two corners. If we consider for example the TWSTFT result for the clock pair Sr(NPL) -- Sr(PTB), with $\bar y^\mathrm{(Sr(NPL)-Sr(PTB))}=-2.9 \times10^{-16}$, the according data point is drawn at $2.9 \times10^{-16}$ on the NPL -- PTB axis in the direction of the ``PTB'' corner.

For clock pairs of different species, similar considerations hold true, but zero relative frequency deviation difference means that the frequency ratio is referenced to the ratio resulting from CIPM recommended values.

First, we discuss the results with respect to the differences between TWSTFT and GPS PPP satellite links.

All combined comparison uncertainties are equal or smaller than $3.5 \times 10^{-16}$. The GPS PPP links yield smaller 
statistical uncertainties than broadband TWSTFT. This is also true for the combined uncertainties, which are dominated by the statistical uncertainties. The smallest combined uncertainty in the case of the GPS PPP link NPL -- PTB is $1.8 \times 10^{-16}$, which is smaller than for the according TWSTFT link by almost a factor of~2.
The TWSTFT comparisons, however, yield slightly smaller systematic uncertainties. Both satellite link techniques lead to results that are compatible within a confidence level of 68~\%. 

Concerning the results in view of the optical clocks, the relative frequency offsets for the clock pairs Yb$^+$(NPL)--Yb$^+$(PTB) and Sr(LNE-SYRTE)--Sr(PTB) are compatible with zero, independent of the link technique. Furthermore, the results for Sr(LNE-SYRTE) -- Sr(PTB) agree with the result of a comparison between these clocks via optical fibres ($0.47\,\pm\,0.5 \times 10^{-16}$), carried out simultaneously with our measurement campaign \cite{Lisdat2016}.

The Sr(LNE-SYRTE)--Sr(NPL) and Sr(PTB)--Sr(NPL) comparisons both show a positive offset, however with the Sr(LNE-SYRTE)--Sr(NPL) comparison still compatible with zero within the $1\sigma$-uncertainty, while the Sr(LNE-SYRTE)--Sr(PTB) results do not indicate a significant offset. Thus, the observed deviations from zero of the Sr(NPL)-related comparisons may indicate that the Sr(NPL) optical frequency was lower by a few $10^{-16}$ compared to the other Sr clocks. 

This observed discrepancy illustrates the value of coordinated clock comparison experiments. At the time of this campaign, accounting for all known systematic effects, the total estimated systematic uncertainty of NPL(Sr) was $6.8 \times 10^{-17}$, as shown in table 2. However, later investigations performed by means of interleaved self-comparison data indicated another uncharacterised systematic frequency shift: depending on the delay between spin-preparation and clock interrogation, the clock frequency could vary by up to $4 \times 10^{-16}$. The working theory is that this could have been a Doppler shift due to radial motion in the lattice, but unfortunately its magnitude at the time of the campaign cannot be quantified in retrospect as the experimental setup has subsequently been modified. The shift has now been eliminated by replacing the lattice laser delivery optics, which has greatly improved the spatial beam quality.

\begin{table*}

\caption{\label{tab:results_odd} Results for the comparisons of optical clocks of different type. The third column gives the optical frequency ratio as an offset from the exact ratio $r_0=1.495\,991\,618\,544\,900$, which is approximately $1.60 \times 10^{-16}$ smaller than the ratio determined from the 2017 CIPM recommended values. The fourth column contains the average relative frequency deviation differences referenced to the 2017 CIPM recommended frequency values. $u_\mathrm{B,c}$ is the combined systematic clock uncertainty, $u_\mathrm{A,l}$ and $u_\mathrm{B,l}$ correspond to the statistical and systematic link uncertainties, $u$ is the combined uncertainty, and $\Delta t_\mathrm{OC}$ corresponds to the effective length of the optical clock comparison. All values except for the last column $\Delta t_\mathrm{OC}$ are in $10^{-16}$.} 
\begin{center}
\footnotesize
\lineup

\begin{tabular}{@{}ccccccccc}

\br
Clock pair & link & $\frac{\nu_\mathrm{\mathrm{Yb}^+}}{\nu_\mathrm{Sr}}-r_0$& $y_\mathrm{Sr}-y_\mathrm{Yb^+}$	& $u_\mathrm{B,c}$	& $u_\mathrm{A,l}$	& $u_\mathrm{B,l}$ & $u$ & $\Delta t_\mathrm{OC}$ [d]\\ \ms
\mr

\multirow{2}{2.3cm}{\centering Sr(LNE-SYRTE) \linebreak-- Yb$^+$(NPL)} 
& \multirow{2}{1.8cm}{\centering TWSTFT\linebreak GPS PPP}
& \multirow{2}{.6cm}{\centering  9.76 \linebreak  11.13}	
& \multirow{2}{.6cm}{\centering -4.9\linebreak -5.8} 
& \multirow{2}{.6cm}{\centering 1.2}	
& \multirow{2}{.6cm}{\centering 3.0\linebreak 2.3}	
& \multirow{2}{.6cm}{\centering 0.7\linebreak 0.8}	
& \multirow{2}{.6cm}{\centering 3.3\linebreak 2.7}
& \multirow{2}{.6cm}{\centering 16.6\linebreak 13.5} \\
\rule[-2ex]{0pt}{8ex}\\

\multirow{2}{2.3cm}{\centering Sr(LNE-SYRTE) \linebreak-- Yb$^+$(PTB)} 
& \multirow{2}{1.8cm}{\centering TWSTFT\linebreak GPS PPP} 
& \multirow{2}{.6cm}{\centering 4.59 \linebreak 8.40}	
& \multirow{2}{.6cm}{\centering -1.4\linebreak -3.9} 
& \multirow{2}{.6cm}{\centering 0.4}	
& \multirow{2}{.6cm}{\centering 2.5\linebreak 1.9}	
& \multirow{2}{.6cm}{\centering 0.9\linebreak 1.2}	
& \multirow{2}{.6cm}{\centering 2.7\linebreak 2.3}
& \multirow{2}{.6cm}{\centering 16.4\linebreak 13.5} \\
\rule[-2ex]{0pt}{8ex}\\

\multirow{2}{2.3cm}{\centering Sr(NPL) \linebreak-- Yb$^+$(PTB)} 
& \multirow{2}{1.8cm}{\centering TWSTFT\linebreak GPS PPP} 
& \multirow{2}{.6cm}{\centering 10.54 \linebreak 9.99}	
& \multirow{2}{.6cm}{\centering -5.4\linebreak -5.0} 
& \multirow{2}{.6cm}{\centering 0.7}	
& \multirow{2}{.6cm}{\centering 3.3\linebreak 1.5}	
& \multirow{2}{.6cm}{\centering 0.5\linebreak 0.6}	
& \multirow{2}{.6cm}{\centering 3.4\linebreak 1.8}
& \multirow{2}{.6cm}{\centering 24.4\linebreak 24.5} \\
\rule[-2ex]{0pt}{8ex}\\

\multirow{2}{2.3cm}{\centering Sr(PTB) \linebreak--Yb$^+$(NPL) } 
& \multirow{2}{1.8cm}{\centering TWSTFT\linebreak GPS PPP} 
& \multirow{2}{.6cm}{\centering 6.44 \linebreak 8.58}	
& \multirow{2}{.6cm}{\centering -2.6\linebreak -4.1} 
& \multirow{2}{.6cm}{\centering 1.1}	
& \multirow{2}{.6cm}{\centering 3.3\linebreak 1.5}	
& \multirow{2}{.6cm}{\centering 0.5\linebreak 0.6}	
& \multirow{2}{.6cm}{\centering 3.5\linebreak 2.0}
& \multirow{2}{.6cm}{\centering 24.4\linebreak 24.5} \\
\rule[-1ex]{0pt}{4ex}\\

\br
\end{tabular}
\end{center}
\end{table*}

\begin{figure*}[]
\centering
\includegraphics[width=160mm]{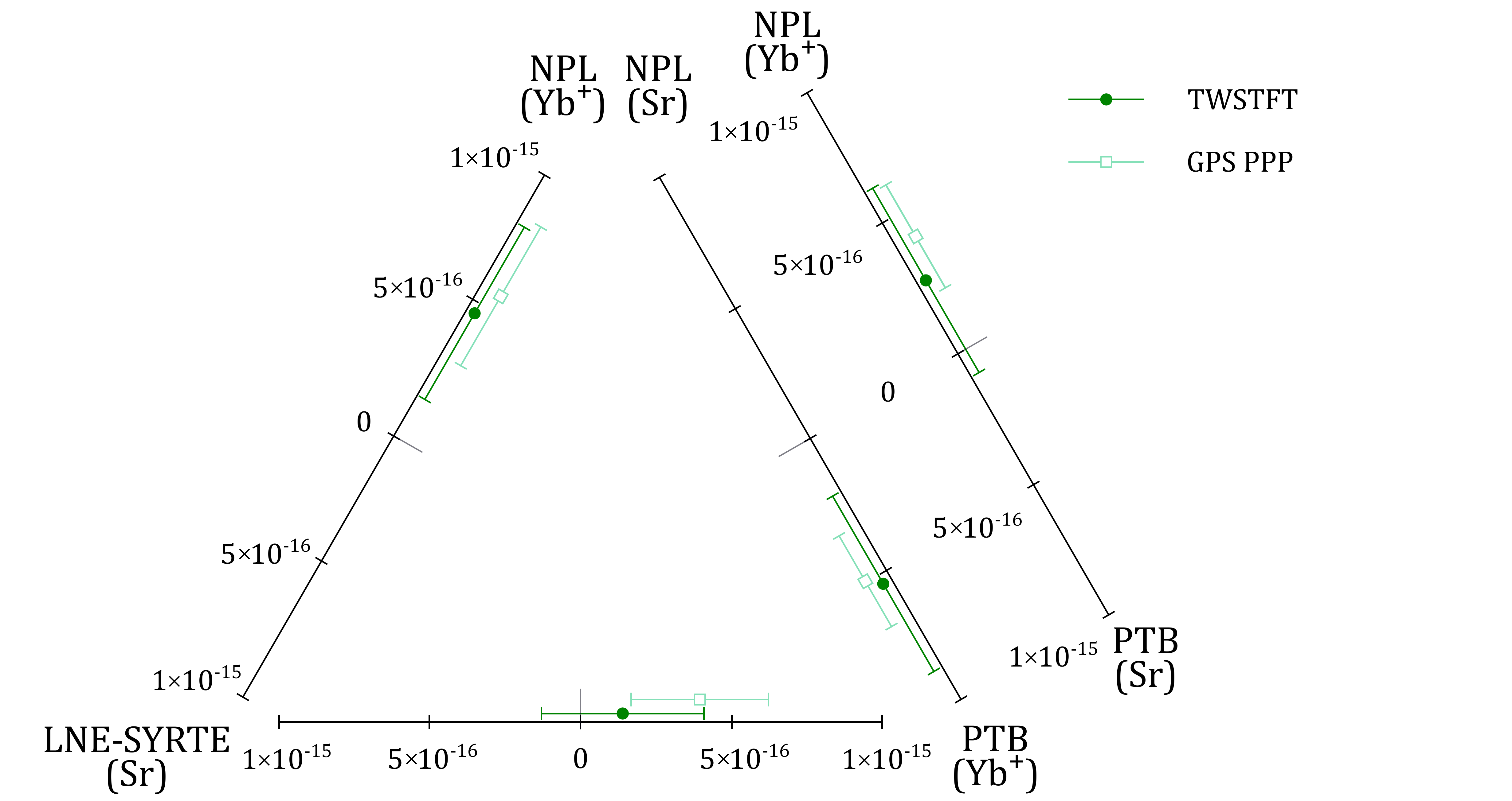}
\caption{The results of the relative frequency deviation differences for the optical clock comparisons between the three institutes LNE-SYRTE, NPL and PTB, for clocks of different type. The triangle plot is explained in the text at the beginning of section~\ref{sec:res}. All results refer to the 2017 CIPM recommended values for the absolute frequencies of the clock transitions for Yb$^+$ and Sr(lattice).}
\label{fig:results_odd}
\end{figure*}

The relative frequency deviation differences for the clocks of different types (Sr lattice against Yb ion) are shown in table~\ref{tab:results_odd}. They are presented relative to the 2017 CIPM recommended values  for the absolute frequencies of the Yb$^+$ octupole transition, $\nu_{\mathrm{Yb_{oct}^+}}=642\,121\,496\,772\,645.0$~Hz and for the Sr transition $\nu_\mathrm{Sr}=429\,228\,004\,229\,873.0$~Hz~\cite{Riehle2018}.
The results are visualized in fig.~\ref{fig:results_odd}, in a similar way as in fig.~\ref{fig:results_even}.
All results deviate from zero in a symmetric way, i.e. $y$(Sr -- Yb$^+$) $<$ 0 for all comparisons, which indicates that the true frequency ratio ${\nu_{\mathrm{Yb}^+}}/{\nu_\mathrm{Sr}}$ is different by a few $10^{-16}$ from the one resulting from the CIPM recommended frequency values.
Nevertheless, it remains compatible with the CIPM recommended values within their combined uncertainty of $7.2 \times 10^{-16}$~\cite{Riehle2018}.


\subsection{Fountain clock comparisons}\label{subsec:res_fountains}
\begin{figure*}[]
\centering
\includegraphics[width=140mm]{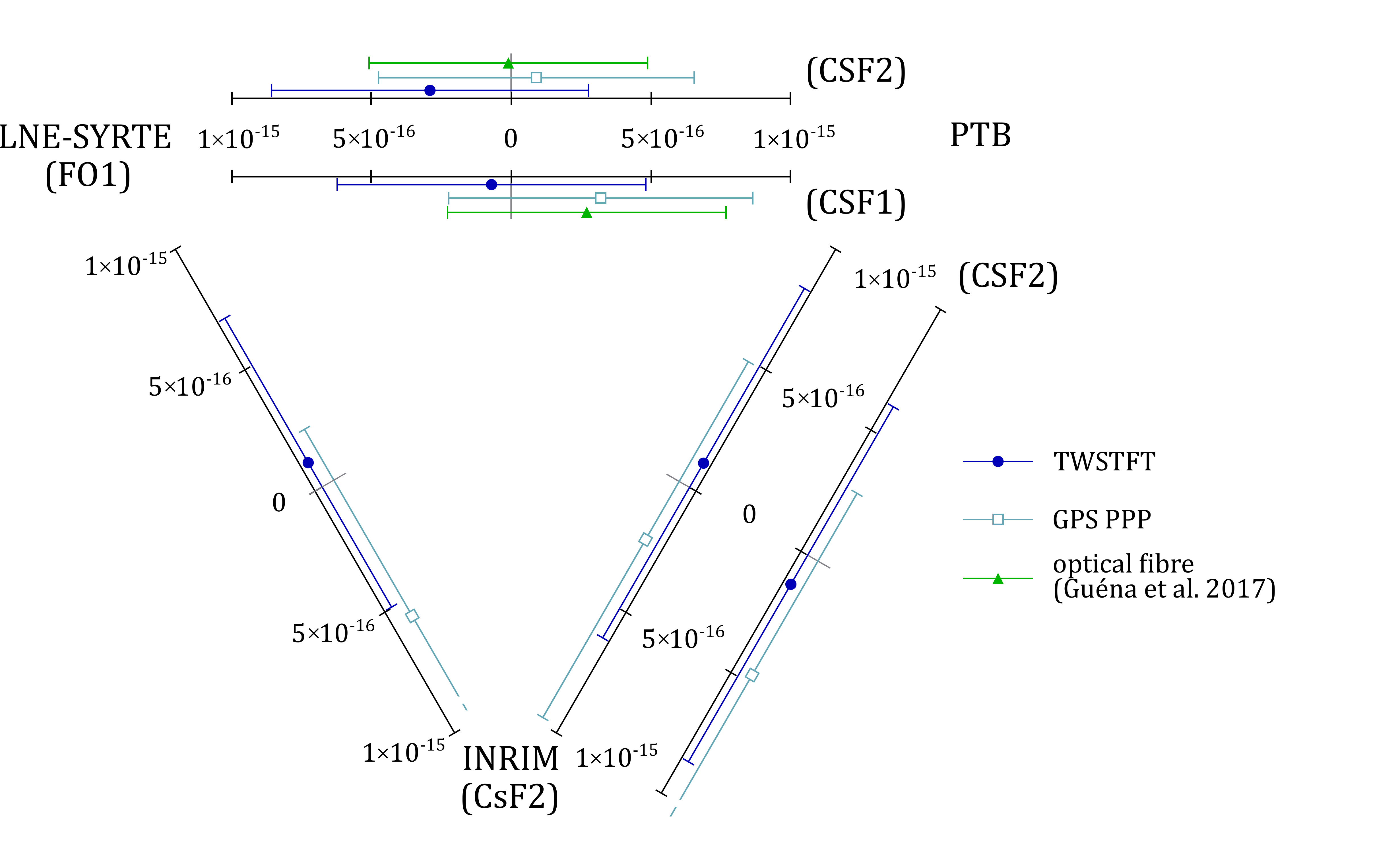}
\caption{The results of the relative frequency deviation differences for all fountain clock comparisons between INRIM and PTB, and all comparisons to FO1 of LNE-SYRTE. The triangle plot is explained in the text at the beginning of section~\ref{sec:res}. The results for the fibre comparison are taken from~\cite{Guena2017}.}
\label{fig:results_csf_fo1}
\end{figure*}

\begin{figure*}[]
\centering
\includegraphics[width=140mm]{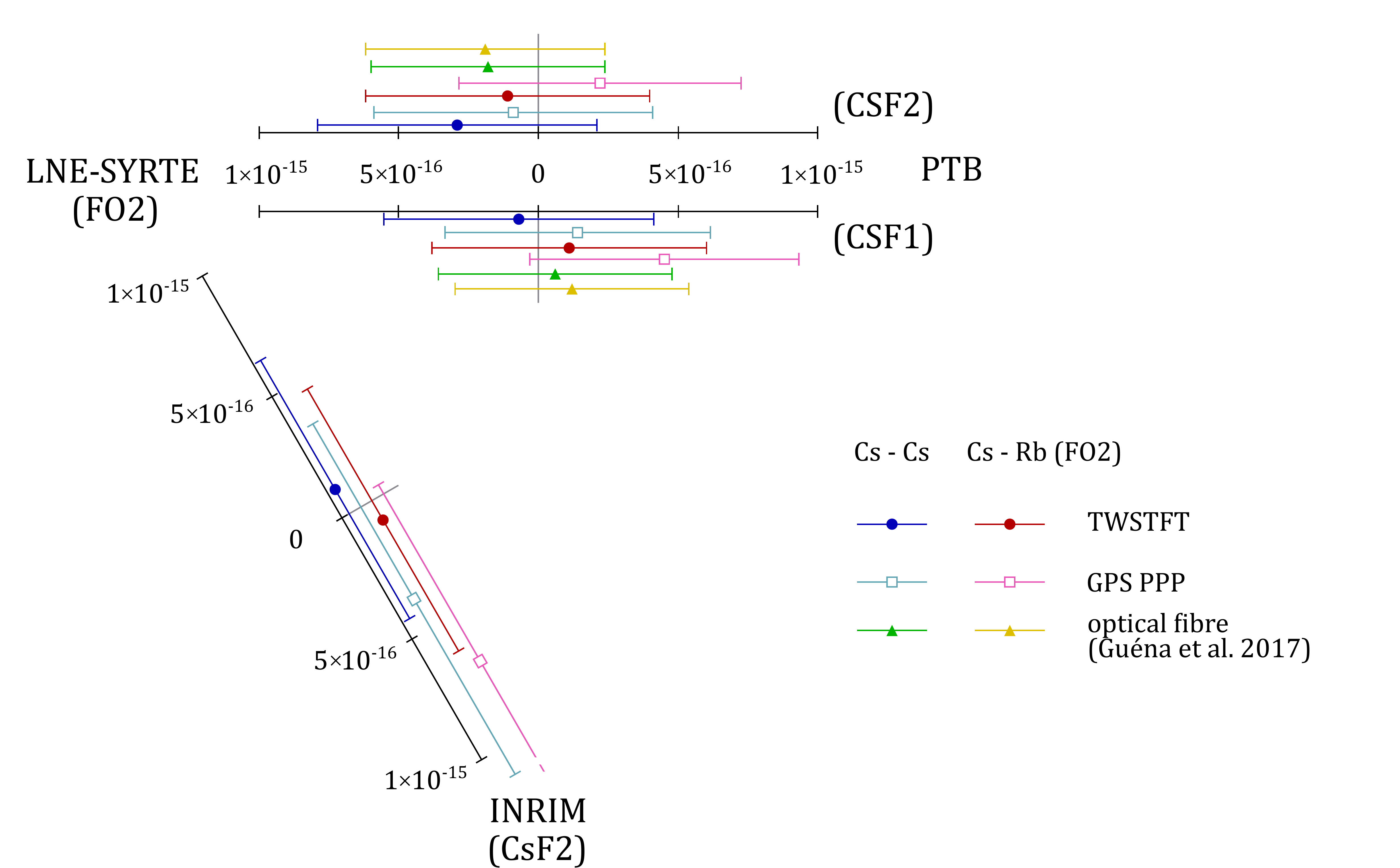}
\caption{The results of the relative frequency deviation differences for all fountain clock comparisons between FO2 (both Cs and Rb) of LNE-SYRTE and the Cs fountains of INRIM and PTB. The results for the fibre comparison are taken from~\cite{Guena2017}. The triangle plot is explained in the text at the beginning of section~\ref{sec:res}. The third side of the triangle, which corresponds to the comparisons between INRIM and PTB, is left out, the results can be found in fig.~\ref{fig:results_csf_fo1}.}
\label{fig:results_csf_fo2}
\end{figure*}

The results of the comparisons between the remote fountain clocks are compiled in table~\ref{tab:FO} and depicted in fig.~\ref{fig:results_csf_fo1} and \ref{fig:results_csf_fo2}. We find good agreement between all six participating fountain clocks at the low $10^{-16}$ level, compatible with the overall comparison uncertainties. We also observe a good agreement between TWSTFT and GPS PPP satellite link techniques. The results for the comparisons between LNE-SYRTE and PTB agree with the frequency differences measured via optical fibre~\cite{Guena2017}.

\begin{table*}

\caption{\label{tab:FO} Summary of the remote fountain clock comparisons. The third column gives the average fountain frequency difference, $u_\mathrm{A,c}$ and $u_\mathrm{B,c}$ are the combined statistical and systematic clock uncertainties (with the dead-time uncertainty as part of the statistical uncertainty), $u_\mathrm{A,l}$ and $u_\mathrm{B,l}$ correspond to the statistical and systematic link uncertainties, and $u$ is the combined uncertainty. The value of the absolute frequency of the $^{87}$Rb ground state hyperfine transition used in these comparisons is the CIPM recommended value 6 834 682 610.904 312 6 Hz of 2017~\cite{Riehle2018}. All values in $10^{-16}$.} 
\begin{center}
\footnotesize
\lineup

\begin{tabular}{@{}cccccccc}

\br
Clock pair & link & difference	& $u_\mathrm{A,c}$	& $u_\mathrm{B,c}$	& $u_\mathrm{A,l}$	& $u_\mathrm{B,l}$ & $u$ \\ \ms
\mr

\multirow{2}{3.8cm}{\centering ITCsF2 (INRIM) \linebreak- FO1 (LNE-SYRTE)} 
& \multirow{2}{1.8cm}{\centering TWSTFT\linebreak GPS PPP} 
& \multirow{2}{.8cm}{\centering -1.0\linebreak \- 5.6\linebreak} 
& \multirow{2}{.8cm}{\centering 2.8\linebreak 3.2}	
& \multirow{2}{.8cm}{\centering 4.3}	
& \multirow{2}{.8cm}{\centering 3.2\linebreak 5.6}	
& \multirow{2}{.8cm}{\centering 0.4\linebreak 0.6} 
& \multirow{2}{.8cm}{\centering 6.0\linebreak 7.7} \\
\rule[-2ex]{0pt}{8ex}\\

\multirow{2}{3.8cm}{\centering ITCsF2 (INRIM) \linebreak- FO2 (LNE-SYRTE)} 
& \multirow{2}{1.8cm}{\centering TWSTFT\linebreak GPS PPP} 
& \multirow{2}{.8cm}{\centering -1.0\linebreak \- 3.8} 
& \multirow{2}{.8cm}{\centering 2.8\linebreak 3.2}	
& \multirow{2}{.8cm}{\centering 3.4}	
& \multirow{2}{.8cm}{\centering 3.2\linebreak 5.6}	
& \multirow{2}{.8cm}{\centering 0.4\linebreak 0.6} 
& \multirow{2}{.8cm}{\centering 5.4\linebreak 7.3}  \\
\rule[-2ex]{0pt}{8ex}\\

\multirow{2}{3.8cm}{\centering ITCsF2 (INRIM) \linebreak- FO2Rb (LNE-SYRTE)} 
& \multirow{2}{1.8cm}{\centering TWSTFT\linebreak GPS PPP} 
& \multirow{2}{.8cm}{\centering \- 0.8\linebreak \- 6.9}   
& \multirow{2}{.8cm}{\centering 2.7\linebreak 3.1}	
& \multirow{2}{.8cm}{\centering 3.5}	
& \multirow{2}{.8cm}{\centering 3.2\linebreak 5.6}	
& \multirow{2}{.8cm}{\centering 0.4\linebreak 0.6} 
& \multirow{2}{.8cm}{\centering 5.4\linebreak 7.3}  \\
\rule[-2ex]{0pt}{8ex}\\

\multirow{2}{3.8cm}{\centering ITCsF2 (INRIM) \linebreak- CSF1 (PTB)}
& \multirow{2}{1.8cm}{\centering TWSTFT\linebreak GPS PPP} 
& \multirow{2}{.8cm}{\centering -1.0\linebreak \- 2.4} 
& \multirow{2}{.8cm}{\centering 2.9\linebreak 3.3}	
& \multirow{2}{.8cm}{\centering 3.8}	
& \multirow{2}{.8cm}{\centering 5.5\linebreak 5.4}	
& \multirow{2}{.8cm}{\centering 0.3\linebreak 0.9} 
& \multirow{2}{.8cm}{\centering 7.2\linebreak 7.4}  \\
\rule[-2ex]{0pt}{8ex}\\

\multirow{2}{3.8cm}{\centering ITCsF2 (INRIM) \linebreak- CSF2 (PTB)}
& \multirow{2}{1.8cm}{\centering TWSTFT\linebreak GPS PPP} 
& \multirow{2}{.8cm}{\centering \- 1.2\linebreak \- 4.7} 
& \multirow{2}{.8cm}{\centering 3.3\linebreak 3.8}	
& \multirow{2}{.8cm}{\centering 3.8}	
& \multirow{2}{.8cm}{\centering 5.5\linebreak 5.4}	
& \multirow{2}{.8cm}{\centering 0.3\linebreak 0.9} 
& \multirow{2}{.8cm}{\centering 7.3\linebreak 7.5} \\
\rule[-2ex]{0pt}{8ex}\\

\multirow{2}{3.8cm}{\centering FO1 (LNE-SYRTE) \linebreak- CSF1 (PTB)}	
& \multirow{2}{1.8cm}{\centering TWSTFT\linebreak GPS PPP} 
& \multirow{2}{.8cm}{\centering \- 0.7\linebreak -3.2} 
& \multirow{2}{.8cm}{\centering 1.4\linebreak 1.7}	
& \multirow{2}{.8cm}{\centering 4.7}	
& \multirow{2}{.8cm}{\centering 2.6\linebreak 2.0}	
& \multirow{2}{.8cm}{\centering 0.3\linebreak 1.0} 
& \multirow{2}{.8cm}{\centering 5.5\linebreak 5.5}  \\
\rule[-2ex]{0pt}{8ex}\\

\multirow{2}{3.8cm}{\centering FO1 (LNE-SYRTE) \linebreak- CSF2 (PTB)} 
& \multirow{2}{1.8cm}{\centering TWSTFT\linebreak GPS PPP} 
& \multirow{2}{.8cm}{\centering \- 2.9\linebreak -0.9} 
& \multirow{2}{.8cm}{\centering 2.1\linebreak 2.5}	
& \multirow{2}{.8cm}{\centering 4.7}	
& \multirow{2}{.8cm}{\centering 2.6\linebreak 2.0}	
& \multirow{2}{.8cm}{\centering 0.3\linebreak 1.0} 
& \multirow{2}{.8cm}{\centering 5.8\linebreak 5.7}  \\
\rule[-2ex]{0pt}{8ex}\\

\multirow{2}{3.8cm}{\centering FO2 (LNE-SYRTE) \linebreak- CSF1 (PTB)} 	
& \multirow{2}{1.8cm}{\centering TWSTFT\linebreak GPS PPP} 
& \multirow{2}{.8cm}{\centering \- 0.7\linebreak -1.4} 
& \multirow{2}{.8cm}{\centering 1.3\linebreak 1.5}	
& \multirow{2}{.8cm}{\centering 3.9}	
& \multirow{2}{.8cm}{\centering 2.6\linebreak 2.0}	
& \multirow{2}{.8cm}{\centering 0.3\linebreak 1.0} 
& \multirow{2}{.8cm}{\centering 4.9\linebreak 4.8}  \\
\rule[-2ex]{0pt}{8ex}\\

\multirow{2}{3.8cm}{\centering FO2 (LNE-SYRTE) \linebreak- CSF2 (PTB)}
& \multirow{2}{1.8cm}{\centering TWSTFT\linebreak GPS PPP} 
& \multirow{2}{.8cm}{\centering \- 2.9\linebreak \- 0.9} 
& \multirow{2}{.8cm}{\centering 2.0\linebreak 2.4}	
& \multirow{2}{.8cm}{\centering 3.9}	
& \multirow{2}{.8cm}{\centering 2.6\linebreak 2.0}	
& \multirow{2}{.8cm}{\centering 0.3\linebreak 1.0} 
& \multirow{2}{.8cm}{\centering 5.1\linebreak 5.0}  \\
\rule[-2ex]{0pt}{8ex}\\

\multirow{2}{3.8cm}{\centering FO2Rb (LNE-SYRTE) \linebreak- CSF1 (PTB)}
& \multirow{2}{1.8cm}{\centering TWSTFT\linebreak GPS PPP} 
& \multirow{2}{.8cm}{\centering -1.1\linebreak -4.5}
& \multirow{2}{.8cm}{\centering 1.2\linebreak 1.4}	
& \multirow{2}{.8cm}{\centering 4.0}	
& \multirow{2}{.8cm}{\centering 2.6\linebreak 2.0}	
& \multirow{2}{.8cm}{\centering 0.3\linebreak 1.0} 
& \multirow{2}{.8cm}{\centering 4.9\linebreak 4.8}  \\
\rule[-2ex]{0pt}{8ex}\\

\multirow{2}{3.8cm}{\centering FO2Rb (LNE-SYRTE) \linebreak- CSF2 (PTB)}
& \multirow{2}{1.8cm}{\centering TWSTFT\linebreak GPS PPP} 
& \multirow{2}{.8cm}{\centering \- 1.1 \linebreak -2.2} 
& \multirow{2}{.8cm}{\centering 2.0\linebreak 2.3}	
& \multirow{2}{.8cm}{\centering 4.0}	
& \multirow{2}{.8cm}{\centering 2.6\linebreak 2.0}	
& \multirow{2}{.8cm}{\centering 0.3\linebreak 1.0} 
& \multirow{2}{.8cm}{\centering 5.2\linebreak 5.1} \\
\rule[-1ex]{0pt}{4ex}\\
\br
\end{tabular}
\end{center}
\end{table*}

\section{Conclusion} \label{sec:con}

We have carried out simultaneous frequency comparisons between remote optical and fountain clocks via satellite-based techniques and reached uncertainties between $1.8 \times 10^{-16}$ and $3.5 \times 10^{-16}$ for the optical clocks and between $4.8 \times 10^{-16}$ and $7.7 \times 10^{-16}$ for the fountain clocks. For this campaign, five optical clocks and six fountain clocks located at four European metrology institutes, INRIM, LNE-SYRTE, NPL and PTB, were compared over a measurement time of several weeks with duty cycles of up to $77\,\%$ for the optical clocks and $98\,\%$ for the fountain clocks. The operated optical clocks were the Sr lattice clocks at NPL, PTB and LNE-SYRTE, and the Yb$^+$ (E3) clocks at NPL and PTB.

The results obtained with the different satellite link techniques are compatible with each other within the combined standard uncertainty. The relative frequency deviation differences between the fountains are consistent with zero within the combined standard uncertainty. The results for comparisons of Sr lattice and Cs/Rb fountain clocks between LNE-SYRTE and PTB agree with comparisons carried out via optical fibre simultaneously with this campaign.

The uncertainties obtained mark a significant improvement for clock comparisons via satellite--based techniques~\cite{Bauch2005, Hachisu2014, Leute2016}.
The statistical uncertainty of the link dominates the overall link uncertainty for all clock comparisons, with GPS PPP yielding lower statistical and thus combined uncertainties than broadband TWSTFT for all optical clock and most fountain clock comparisons. The TWSTFT measurements suffered from unforeseen technical disturbances causing gaps and an increased instability for averaging times larger than 1\,day. GPS measurements on the other hand were uninterrupted, and apart from a few limitations for the receivers of LNE-SYRTE and INRIM, data over the whole campaign could be used.

To achieve even lower uncertainties, further characterization and improvement of the robustness of stations and links is needed, especially in the TWSTFT case.  
Particularly, software defined radio (SDR) receivers as a replacement for the analog delay locked loops of the SATRE modems’ Rx modules demonstrated a significant reduction of instabilities in TWSTFT links contributing to the production of UTC~\cite{Jiang_2018}. SDR receivers are significantly less sensitive to interferences between several signals transmitted through a single transponder as used in this comparison~\cite{Jiang_2019}.

In summary, the two exploited satellite link techniques are complementary as they may be considered as completely independent. This offers the possibility to find undiscovered or study obscured systematic effects in both techniques limiting the significance of frequency comparisons especially for those long-baseline links were optical fiber links do not exist.

We have shown an example of how data sets with different noise types, gaps and correlations can be treated while avoiding unnecessarily discarding data and being limited by short-term noise processes. 

This first large-scale coordinated international clock comparison campaign indicated a possible unexpected systematic frequency shift in NPL's Sr clock. Further investigations motivated by this result have led to an understanding and significant suppression of a systematic frequency shift related to radial motion in the lattice.

This work provides satisfactory tests of reliability between clocks of the same species, with agreement matching the uncertainty of the current definition of the SI second: it is an essential step towards a possible redefinition of the second. The clock comparisons pushed towards the ascertainting the ratio between different species, which is useful in the perspective of standards that will be secondary representations after a redefinition of the second. Finally, the enhanced microwave satellite techniques implemented, despite the fact that they remain behind the optical fiber links for the frequency comparisons, allow these comparisons to the very low $10^{-16}$ level.

\section*{Acknowledgment}
This work was carried out within the framework of the project \textit{SIB55 ITOC, International Timescales with Optical Clocks}, a part of the European Metrology Research Programme EMRP. The EMRP is jointly funded by the EMRP participating countries within EURAMET and the European Union. The authors thank W. Sch\"afer from TimeTech for stimulating discussions and support, and the Institute of National Resources Canada for providing the software license of their PPP calculation software. The authors thank S\'ebastien Merlet for his support to gravimetric and levelling measurements at and around the LNE-SYRTE site.

\section*{Appendix} \label{sec:app}
\subsection*{Estimation of autocovariances and of the standard deviation of the mean} \label{sec:ex1}
In this appendix we describe how the mean frequency value and its statistical uncertainty, i.e. the standard deviation of the mean, are determined from the frequency data. The frequency values are serially correlated and are given on a time grid with regular spacing (e.g. 1\,s), however with some of the frequency values missing or invalid at gaps. Text-book statistical procedures for the determination of the standard deviation of the mean either do not take into account serial correlations or do not allow for gaps. They can only be used to determine conservative estimates for the standard deviation of the mean, which are not useful in our case. For this reason, we have derived a procedure which takes all these aspects into account. As the fundamental prerequisite for the concept of representing a measurement result by a mean value and its statistical uncertainty, ergodicity of the statistical processes is assumed as usual.

The frequency data given on a regular time grid can be represented by a uniform discrete time series of frequency values $y_i$, $i=1 \dots N$. For the treatment of the gaps, a time series of the same length $N$ comprising associated weights $w_i$ is introduced, which are set to zero at times of missing or invalid frequency data. The weights $w_i$ used are not all zero and not fully random, sometimes referred to as reliability weights. This allows us to derive an estimate of the mean and its standard deviation from the frequency data.

Independently of the correlations, the unbiased estimate for the weighted mean is given by the weighted sample mean
\begin{equation}
\bar{y}_w=\frac{\sum_{i=1}^N w_i y_i}{\sum_{i=1}^N w_i}.
\label{eq:mean}
\end{equation}

\noindent{In general, the variance of a random variable $X$ is defined as}
\begin{eqnarray}
\sigma_X^2&=&\mathbb{E}\left[\left(X-\mathbb{E}[X]\right)^2\right]=\mathbb{E}\left[\left(X-\mu\right)^2\right]\nonumber\\
&=&\mathbb{E}\left[X^2\right]-\mu^2,
\label{eq:variance}
\end{eqnarray}
\noindent{with $\mathbb{E}[\ldots]$ denoting the expected value of the random variable in the square brackets, and with $\mu$ = $\mathbb{E}[X]$.}
In order to estimate the population variance $\sigma^2$ and the standard deviation of the weighted sample mean $\sigma_{\bar{y}_w}$, we must take both the correlations and the weighting into account. 

It is conducive to start with the estimation of the autocovariances. They carry the information about the serial correlations and thus play an essential role in the derivation of the estimators for the population variance and for the variance of the mean. For a stationary process such as any ergodic process, the lag-$l$ autocovariance is defined as 
\begin{eqnarray}
R_l&=&\mathbb{E}\left[\left(y_i-\mu\right)\left(y_{i+l}-\mu\right)\right]\nonumber\\
&=&\mathbb{E}\left[\left(y_i y_{i+l}\right)\right] - \mu^2.
\label{eq:R_l}
\end{eqnarray}
\noindent{For lag $l=0$, this is the population variance}
\begin{eqnarray}
R_0&=&\sigma^2.
\label{eq:R_0}
\end{eqnarray}
\noindent{An estimator for the lag-$l$ autocovariance is the weighted sample autocovariance}
\begin{eqnarray}
\hat{R}_l&=&\frac{\sum_{i=1}^{N-l}\sqrt{w_i w_{i+l}}\left(y_i-\bar{y}_w\right)\left(y_{i+l}-\bar{y}_w\right)}{\sum_{i=1}^{N-l}\sqrt{w_i w_{i+l}}}.
\label{eq:Rhat_l}
\end{eqnarray}
\noindent{We will now derive its expected value. First, note the relation}
\begin{eqnarray}
\mathbb{E}\left[\bar{y}_w^2\right] &=& \sum_{i=1}^N w_i\sum_{j=1}^N  w_j\mathbb{E}\left[y_i y_j\right]\nonumber\\
&=&   \mathbb{E}\left[y_i \sum_{j=1}^N w_j y_j\right]\nonumber\\
&=&\mathbb{E}\left[y_i \bar{y}_w\right].
\label{eq:E_y_yw_app}
\end{eqnarray}
\noindent{With this at hand, we can show for the expected value of the weighted sample autocovariance:}
\begin{eqnarray}
\mathbb{E}\left[\hat{R}_l\right]&=&\frac{\sum_{i=1}^{N-l}\sqrt{w_i w_{i+l}}\,\mathbb{E}\left[\left(y_i-\bar{y}_w\right)\left(y_{i+l}-\bar{y}_w\right)\right]}{\sum_{i=1}^{N-l}\sqrt{w_i w_{i+l}}}\nonumber\\
&=&\mathbb{E}\left[\left(y_i-\bar{y}_w\right)\left(y_{i+l}-\bar{y}_w\right)\right]\nonumber\\
&=&\mathbb{E}\left[y_i y_{i+l}\right]-\mu^2
-\mathbb{E}\left[y_i \bar{y}_w\right]
-\mathbb{E}\left[y_{i+l} \bar{y}_w\right]\nonumber\\
& &+\mathbb{E}\left[\bar{y}_w^2\right]+\mu^2\nonumber\\
&=&\left(\mathbb{E}\left[y_i y_{i+l}\right]-\mu^2\right)
-\left(\mathbb{E}\left[\bar{y}_w^2\right]-\mu^2\right)\nonumber\\
&=&R_l-\sigma_{\bar{y}_w}^2
\label{eq:E_Rhat_l}.
\end{eqnarray}
\noindent{Since the variance of the mean is always larger than zero, the estimator $\hat{R}_l$ is negatively biased. To the best of our knowledge no unbiased estimator of the autocovariance has been proposed yet and it probably does not exist, but the sample covariance estimator is sometimes denoted as ``unbiased estimator'' in statistics textbooks even though it is actually biased~\cite{Percival1993}. For zero lag we get the relation}
\begin{eqnarray}
\mathbb{E}\left[\hat{R}_0\right]&=&R_0-\sigma_{\bar{y}_w}^2.
\label{eq:E_Rhat_0}
\end{eqnarray}
\noindent{For the derivation of the variance of the weighted mean first note that}
\begin{eqnarray}
\bar{y}_w^2&=&\sum_{i=1}^{N}\sum_{j=1}^{N}w_i w_j y_i y_j\nonumber\\
&=& \sum_{i=1}^{N} w_i^2 y_i^2 +2\sum_{i=1}^{N-1}\sum_{j=1}^{N-i}w_j w_{i+j} y_j y_{i+j},\nonumber\\
\label{eq:y_w_app}
\end{eqnarray}
\noindent{and its expected value}
\begin{eqnarray}
\mathbb{E}\left[\bar{y}_w^2\right]&=& \sum_{i=1}^{N} w_i^2 \mathbb{E}\left[y_i^2\right]+2\sum_{i=1}^{N-1}\sum_{j=1}^{N-i}w_j w_{i+j} \mathbb{E}\left[y_j y_{i+j}\right]\nonumber\\
&=& \sum_{i=1}^{N} w_i^2 \left(R_0+\mu^2\right) \nonumber\\
& & +2\sum_{i=1}^{N-1}\sum_{j=1}^{N-i}w_j w_{i+j} \left(R_i+\mu^2\right) \nonumber\\
&=& \sum_{i=1}^{N} w_i^2 R_0 +2\sum_{i=1}^{N-1}R_i\sum_{j=1}^{N-i}w_j w_{i+j} \nonumber\\
& & +\left(\sum_{i=1}^{N} w_i^2+2\sum_{i=1}^{N-1}\sum_{j=1}^{N-i}w_j w_{i+j}\right)\mu^2\nonumber\\
&=& \sum_{i=1}^{N} w_i^2 R_0 +2\sum_{i=1}^{N-1}R_i\sum_{j=1}^{N-i}w_j w_{i+j} \nonumber\\
& & +\left(\sum_{i=1}^{N}\sum_{j=1}^{N}w_i w_j\right)\mu^2\nonumber\\
&=& \sum_{i=1}^{N} w_i^2 R_0 +2\sum_{i=1}^{N-1}R_i\sum_{j=1}^{N-i}w_j w_{i+j} +\mu^2. 
\label{eq:E_y_w_app} 
\end{eqnarray}
\noindent{Hence, we get for the variance of the weighted mean}
\begin{eqnarray}
\sigma_{\bar{y}_w}^2&=&\mathbb{E}\left[\bar{y}_w^2\right]-\mu^2\nonumber\\
&=&R_0 s_0 + 2\sum_{i=1}^{N-1}R_i s_i \nonumber\\
&=&r R_0
\label{eq:r_app}
\end{eqnarray}
\noindent{with}
\begin{eqnarray}
s_i&=&\sum_{j=1}^{N-i} w_j w_{i+j}\quad \textrm{for} \quad i = 0\dots N-1
\label{eq:s_i}
\end{eqnarray}
\noindent{and}
\begin{eqnarray}
r&=&s_0 + 2\sum_{i=1}^{N-1}\frac{R_i}{R_0} s_i.
\label{eq:r}
\end{eqnarray}
\noindent{If all weighting factors are equal, $w_i=1/N$, it follows that $s_i=(N-i)/N$ in accordance with the formulae in~\cite{Zhang2006, Zieba2010}.}
The factor $r$ itself depends on all autocovariances $R_l$ and has been reported previously~\cite{Zhang2006}, though without a derivation.
Equation~\ref{eq:r_app} is the generalization of the well-known relation
\begin{equation}
\sigma_{\bar{y}_w}^2 = \sigma^2/N,
\label{eq:sigma_mean_uncorrelated}
\end{equation}
\noindent{which is only applicable to uncorrelated and unweighted data.} Due to this fact, the inverse of the factor $r$ is often interpreted as an effective number of independent observations ($N_\mathrm{eff}$, or similar nomenclature). Although this is undoubtedly a very descriptive interpretation in the case of an equally weighted mean \cite{Zieba2010}, it at least partially loses its sense and can actually be misleading in the case of nonuniform weights, especially if some of the weighting factors are zero or very small.

Since we do not know the true autocovariances $R_l$, and unbiased estimates of the autocovariances do not exist, the best that can be done in this situation is to follow the approach of references \cite{Zhang2006} and \cite{Zieba2011} and plug the biased estimators $\hat{R}_l$ from eq.~\ref{eq:Rhat_l} into eq.~\ref{eq:r_app}:
\begin{eqnarray}
\hat{\sigma}_{\bar{y}_w}^2&=&\hat{R}_0 s_0 + 2\sum_{i=1}^{N-1}\hat{R}_i s_i.
\label{eq:Rhat}
\end{eqnarray}
\noindent{This results in a biased estimator $\hat{\sigma}_{\bar{y}_w}^2$ for the variance of the mean. However, the bias is reduced with respect to naively using eq.~\ref{eq:sigma_mean_uncorrelated}, because at least approximate information on the correlations and weightings is used.}

When doing so, we have to make sure that the estimators for the autocorrelation coefficient are statistically representative, i.e. that the sums in equation~\ref{eq:Rhat_l} include enough elements. For small lags $l$, this is usually the case, but not for large lags. In many practical cases, we know that the envelope of the autocorrelation asymptotically decreases to zero for large lags. For this reason, the usual approach is to set $\hat{R}_l=0$ for lags $l$ larger than a cutoff-lag $l_{cut}$. Several criteria to determine $l_{cut}$ have been suggested, such as a ``first transit through zero'' (FTZ)-criterion~\cite{Zieba2011}, an $l_{cut}\approx N/4$ rule of thumb, or more sophisticated criteria for specific statistical processes~\cite{Zhang2006}. It can also be noticed from eq.~\ref{eq:Rhat} that setting negative $\hat{R}_l<0$ to zero yields a more conservative estimator for the variance of the mean.

\section*{References}
\bibliographystyle{ieeetr}
\bibliography{ITOC_Paper_Bib}

\begin{thebibliography}{10}

\bibitem{Wynands2005}
R.~Wynands and S.~Weyers, ``{Atomic fountain clocks},'' {\em Metrologia},
  vol.~42, no.~3, pp.~S64--S79, 2005.

\bibitem{Levi2014}
F.~Levi, D.~Calonico, C.~E. Calosso, A.~Godone, S.~Micalizio, and G.~A.
  Costanzo, ``{Accuracy evaluation of ITCsF2: a nitrogen cooled caesium
  fountain},'' {\em Metrologia}, vol.~51, no.~3, pp.~270--284, 2014.

\bibitem{Guena2012a}
J.~Gu{\'e}na, M.~Abgrall, D.~Rovera, P.~Laurent, B.~Chupin, M.~Lours,
  G.~Santarelli, P.~Rosenbusch, M.~E. Tobar, R.~Li, K.~Gibble, A.~Clairon, and
  S.~Bize, ``{Progress in Atomic Fountains at LNE-SYRTE},'' {\em IEEE
  Transactions on Ultrasonics, Ferroelectrics, and Frequency Control}, vol.~59,
  no.~3, pp.~391--410, 2012.

\bibitem{wey18}
S.~Weyers, V.~Gerginov, M.~Kazda, J.~Rahm, B.~Lipphardt, G.~Dobrev, and
  K.~Gibble, ``Advances in the accuracy, stability, and reliability of the
  {PTB} primary fountain clocks,'' {\em Metrologia}, vol.~55, no.~6,
  pp.~789--805, 2018.

\bibitem{nic15}
T.~L. Nicholson, S.~L. Campbell, R.~B. Hutson, G.~E. Marti, B.~J. Bloom, R.~L.
  McNally, W.~Zhang, M.~D. Barrett, M.~S. Safronova, G.~F. Strouse, W.~L. Tew,
  and J.~Ye, ``Systematic evaluation of an atomic clock at $2 \times 10^{-18}$
  total uncertainty,'' {\em Nature Communications}, vol.~6, p.~6896, 2015.

\bibitem{Huntemann2016}
N.~Huntemann, C.~Sanner, B.~Lipphardt, C.~Tamm, and E.~Peik, ``{Single-ion
  atomic clock with $3\times\,10^{-18}$ systematic uncertainty},'' {\em
  Physical Review Letters}, vol.~116, no.~6, p.~063001, 2016.

\bibitem{gre18}
W.~F. McGrew, X.~Zhang, R.~J. Fasano, S.~A. Schäffer, K.~Beloy, D.~Nicolodi,
  R.~C. Brown, N.~Hinkley, G.~Milani, M.~Schioppo, T.~H. Yoon, and A.~D.
  Ludlow, ``Atomic clock performance enabling geodesy below the centimetre
  level,'' {\em Nature}, vol.~564, no.~7734, pp.~87--90, 2018.

\bibitem{bre19}
S.~M. Brewer, J.-S. Chen, A.~M. Hankin, E.~R. Clements, C.~W. Chou, D.~J.
  Wineland, D.~B. Hume, and D.~R. Leibrandt, ``$^{27}${Al}$^{+}$ quantum-logic
  clock with a systematic uncertainty below ${10}^{\ensuremath{-}18}$,'' {\em
  Phys. Rev. Lett.}, vol.~123, p.~033201, 2019.

\bibitem{Riehle2015}
F.~Riehle, ``{Towards a redefinition of the second based on optical atomic
  clocks},'' {\em Comptes Rendus Physique}, vol.~16, no.~5, pp.~506--515, 2015.

\bibitem{Gill2016}
P.~Gill, ``{Is the time right for a redefinition of the second by optical
  atomic clocks?},'' in {\em Journal of Physics: Conference Series}, vol.~723,
  p.~012053, IOP Publishing, 2016.

\bibitem{biz19}
S.~Bize, ``The unit of time: Present and future directions,'' {\em Comptes
  Rendus Physique}, vol.~20, pp.~153--168, 2019.

\bibitem{Parker2001}
T.~Parker, P.~Hetzel, S.~Jefferts, S.~Weyers, L.~Nelson, A.~Bauch, and
  J.~Levine, ``{First comparison of remote cesium fountains},'' in {\em
  Proceedings of the 2001 IEEE International Frequency Control Symposium and
  PDA Exhibition.}, pp.~63--68, IEEE, 2001.

\bibitem{Bauch2005}
A.~Bauch, J.~Achkar, S.~Bize, D.~Calonico, R.~Dach, R.~Hlava{\'c}, L.~Lorini,
  T.~Parker, G.~Petit, and D.~Piester, ``{Comparison between frequency
  standards in Europe and the USA at the $10^{-15}$ uncertainty level},'' {\em
  Metrologia}, vol.~43, no.~1, pp.~109--120, 2005.

\bibitem{Fujieda2007}
M.~Fujieda, T.~Gotoh, D.~Piester, M.~Kumagai, S.~Weyers, A.~Bauch, R.~Wynands,
  and M.~Hosokawa, ``{First comparison of primary frequency standards between
  Europe and Asia},'' in {\em IEEE International Frequency Control Symposium,
  Joint with the 21st European Frequency and Time Forum.}, pp.~937--941, IEEE,
  2007.

\bibitem{Zhang2014}
A.~Zhang, K.~Liang, Z.~Yang, F.~Fang, T.~Li, D.~Piester, V.~Gerginov,
  S.~Weyers, A.~Bauch, M.~Fujieda, I.~Blinov, A.~Boiko, Y.~Domnin, A.~Naumov,
  Y.~Smirnov, A.~S. Gupta, P.~Arora, A.~Acharya, and A.~Agarwal, ``Comparison
  of caesium fountain clocks in {E}urope and {A}sia,'' in {\em European
  Frequency and Time Forum (EFTF)}, pp.~447--450, 2014.

\bibitem{Guena2017}
J.~Gu\'ena, S.~Weyers, M.~Abgrall, C.~Grebing, V.~Gerginov, P.~Rosenbusch,
  S.~Bize, B.~Lipphardt, H.~Denker, N.~Quintin, S.~M.~F. Raupach, D.~Nicolodi,
  F.~Stefani, N.~Chiodo, S.~Koke, A.~Kuhl, F.~Wiotte, F.~Meynadier,
  E.~Camisard, C.~Chardonnet, Y.~{Le Coq}, M.~Lours, G.~Santarelli,
  A.~Amy-Klein, R.~{Le Targat}, O.~Lopez, P.~E. Pottie, and G.~Grosche,
  ``{First international comparison of fountain primary frequency standards via
  a long distance optical fiber link},'' {\em Metrologia}, vol.~54, no.~3,
  pp.~348--354, 2017.

\bibitem{yam11}
A.~Yamaguchi, M.~Fujieda, M.~Kumagai, H.~Hachisu, S.~Nagano, Y.~Li, T.~Ido,
  T.~Takano, M.~Takamoto, and H.~Katori, ``Direct comparison of distant optical
  lattice clocks at the $10^{-16}$ uncertainty,'' {\em Appl. Phys. Express},
  vol.~4, no.~8, p.~082203, 2011.

\bibitem{Hachisu2014}
H.~Hachisu, M.~Fujieda, S.~Nagano, T.~Gotoh, A.~Nogami, T.~Ido, S.~Falke,
  N.~Huntemann, C.~Grebing, B.~Lipphardt, C.~Lisdat, and D.~Piester, ``Direct
  comparison of optical lattice clocks with an intercontinental baseline of 9
  000 km,'' {\em Optics Letters}, vol.~39, pp.~4072--4075, 2014.

\bibitem{tak16}
T.~Takano, M.~Takamoto, I.~Ushijima, N.~Ohmae, T.~Akatsuka, A.~Yamaguchi,
  Y.~Kuroishi, H.~Munekane, B.~Miyahara, and H.~Katori, ``Geopotential
  measurements with synchronously linked optical lattice clocks,'' {\em Nature
  Photonics}, vol.~10, pp.~662--666, 2016.

\bibitem{Lisdat2016}
C.~Lisdat, G.~Grosche, N.~Quintin, C.~Shi, S.~Raupach, C.~Grebing, D.~Nicolodi,
  F.~Stefani, A.~Al-Masoudi, S.~D\"orscher, S.~H\"afner, J.-L. Robyr,
  N.~Chiodo, S.~Bilicki, E.~Bookjans, A.~Koczwara, S.~Koke, A.~Kuhl, F.~Wiotte,
  F.~Meynadier, E.~Camisard, M.~Abgrall, M.~Lours, T.~Legero, H.~Schnatz,
  U.~Sterr, H.~Denker, C.~Chardonnet, Y.~Le~Coq, G.~Santarelli, A.~Amy-Klein,
  R.~Le~Targat, J.~Lodewyck, O.~Lopez, and P.-E. Pottie, ``A clock network for
  geodesy and fundamental science,'' {\em Nature Communications}, vol.~7,
  p.~12443, 2016.

\bibitem{Leute2016}
J.~Leute, N.~Huntemann, B.~Lipphardt, C.~Tamm, P.~B.~R. Nisbet-Jones, S.~A.
  King, R.~M. Godun, J.~M. Jones, H.~S. Margolis, P.~B. Whibberley, A.~Wallin,
  M.~Merimaa, P.~Gill, and E.~Peik, ``Frequency comparison of {$^{171}$Yb$^+$}
  ion optical clocks at {PTB} and {NPL} via {GPS} {PPP},'' {\em IEEE
  Transactions on Ultrasonics, Ferroelectrics, and Frequency Control}, vol.~63,
  no.~7, pp.~981--985, 2016.

\bibitem{gre19}
W.~F. McGrew, X.~Zhang, H.~Leopardi, R.~J. Fasano, D.~Nicolodi, K.~Beloy,
  J.~Yao, J.~A. Sherman, S.~A. Sch\"{a}ffer, J.~Savory, R.~C. Brown,
  S.~R\"{o}misch, C.~W. Oates, T.~E. Parker, T.~M. Fortier, and A.~D. Ludlow,
  ``Towards the optical second: verifying optical clocks at the {SI} limit,''
  {\em Optica}, vol.~6, no.~4, pp.~448--454, 2019.

\bibitem{Piester2008a}
D.~Piester, A.~Bauch, J.~Becker, E.~Staliuniene, and C.~Schlunegger, ``{On
  measurement noise in the European TWSTFT network},'' {\em IEEE Transactions
  on Ultrasonics, Ferroelectrics, and Frequency Control}, vol.~55, no.~9,
  pp.~1906--1912, 2008.

\bibitem{lod16}
J.~Lodewyck, S.~Bilicki, E.~Bookjans, J.-L. Robyr, C.~Shi, G.~Vallet,
  R.~Le~Targat, D.~Nicolodi, Y.~Le~Coq, J.~Gu\'ena, M.~Abgrall, P.~Rosenbusch,
  and S.~Bize, ``Optical to microwave clock frequency ratios with a nearly
  continuous strontium optical lattice clock,'' {\em Metrologia}, vol.~53,
  no.~4, pp.~1123--1130, 2016.

\bibitem{gue11}
J.~Gu{\'e}na, R.~Li, K.~Gibble, S.~Bize, and A.~Clairon, ``Evaluation of
  {Doppler} shifts to improve the accuracy of primary atomic fountain clocks,''
  {\em Phys. Rev. Lett.}, vol.~106, no.~13, p.~130801, 2011.

\bibitem{Guena2014}
J.~Gu\'ena, M.~Abgrall, A.~Clairon, and S.~Bize, ``{Contributing to TAI with a
  secondary representation of the SI second},'' {\em Metrologia}, vol.~51,
  no.~1, pp.~108--120, 2014.

\bibitem{Baynham2018}
C.~F.~A. Baynham, R.~M. Godun, J.~M. Jones, S.~A. King, P.~B.~R. Nisbet-Jones,
  F.~Baynes, A.~Rolland, P.~E.~G. Baird, K.~Bongs, P.~Gill, and H.~S. Margolis,
  ``Absolute frequency measurement of the optical clock transition in with an
  uncertainty of using a frequency link to international atomic time,'' {\em
  Journal of Modern Optics}, vol.~65, no.~5-6, pp.~585--591, 2018.

\bibitem{comm1}
The relevant fountain systematic uncertainty budgets of CSF1 and CSF2 of PTB
  are close to those reported to the BIPM for the June 2015 evaluations
  (Circular T330~\cite{CircT}). In the case of CSF1, a reevaluation of the
  distributed cavity phase shift was performed after the comparison campaign,
  which retrospectively leads to a significantly reduced overall systematic
  uncertainty.

\bibitem{voi16}
C.~Voigt, H.~Denker, and L.~Timmen, ``Time-variable gravity potential
  components for optical clock comparisons and the definition of international
  time scales,'' {\em Metrologia}, vol.~53, no.~6, p.~1365, 2016.

\bibitem{Denker2018}
H.~Denker, L.~Timmen, C.~Voigt, S.~Weyers, E.~Peik, H.~S. Margolis, P.~Delva,
  P.~Wolf, and G.~Petit, ``Geodetic methods to determine the relativistic
  redshift at the level of $10^{-18}$ in the context of international
  timescales: a review and practical results,'' {\em Journal of Geodesy},
  vol.~92, no.~5, pp.~487--516, 2018.

\bibitem{meh18}
T.~Mehlst{\"a}ubler, G.~Grosche, C.~Lisdat, P.~Schmidt, and H.~Denker, ``Atomic
  clocks for geodesy,'' {\em Rep. Prog. Phys.}, vol.~81, p.~064401, 2018.

\bibitem{del19}
P.~Delva, H.~Denker, and G.~Lion, ``Chronometric {{Geodesy}}: {{Methods}}
  and{{Applications}},'' in {\em Relativistic {{Geodesy}}: {{Foundations}}
  and{{Applications}}} (D.~Puetzfeld and C.~L{\"a}mmerzahl, eds.), Fundamental
  {{Theories}} of {{Physics}}, pp.~25--85, {Cham}: {Springer International
  Publishing}, 2019.

\bibitem{ITU_TW}
``{The operational use of two-way satellite time and frequency transfer
  employing pseudorandom noise codes}.'' Recommendation ITU-R TF. 1153-4, ITU,
  Geneva, Switzerland, 2015.

\bibitem{Petit1994}
G.~Petit and P.~Wolf, ``{Relativistic theory for picosecond time transfer in
  the vicinity of the Earth},'' {\em Astronomy and Astrophysics}, vol.~286,
  pp.~971--977, 1994.

\bibitem{Piester2007}
D.~Piester, A.~Bauch, M.~Fujieda, T.~Gotoh, M.~Aida, H.~Maeno, M.~Hosokawa, and
  S.~Yang, ``{Studies on instabilities in long-baseline two-way satellite time
  and frequency transfer (TWSTFT) including a troposphere delay model},'' in
  {\em Proc. 39th Annual Precise Time and Time Interval (PTTI) Systems and
  Applications Meeting, 27-29 Nov 2007, Long Beach, California, USA},
  pp.~211--222, 2007.

\bibitem{TEC_data}
``{Ionosphere GNSS products}.'' \url{ftp://gnss.oma.be/gnss/products/IONEX/}.
\newblock Retrieved in July 2017.

\bibitem{Xubook2003}
G.~Xu, {\em {GPS - Theory, Algorithms and Applications}}, ch.~Physical
  Influences of GPS Surveying: Ionospheric effects, pp.~39--50.
\newblock Springer-Verlag Berlin Heidelberg New York, 2003.

\bibitem{kou2001}
J.~Kouba and P.~H{\'e}roux, ``Precise point positioning using igs orbit and
  clock products,'' {\em GPS Solutions}, vol.~5, pp.~12--28, 2001.

\bibitem{sch2007}
R.~Schmid, P.~Steigenberger, G.~Gendt, M.~Ge, and M.~Rothacher, ``Generation of
  a consistent absolute phase-center correction model for {GPS} receiver and
  satellite antennas,'' {\em Journal of Geodesy}, vol.~81, pp.~781--798, 2007.

\bibitem{wu1993}
J.~T. {Wu}, S.~C. {Wu}, G.~A. {Hajj}, W.~I. {Bertiger}, and S.~M. {Lichten},
  ``{Effects of antenna orientation on {GPS} carrier phase},'' in {\em
  Astrodynamics 1991} (P.~A. {Penzo} and D.~F. {Bender}, eds.), pp.~1647--1660,
  1992.

\bibitem{ash2003}
N.~Ashby, ``Relativity in the {G}lobal {P}ositioning {S}ystem,'' {\em Living
  Reviews in Relativity}, vol.~6, no.~1, p.~1, 2003.

\bibitem{sul90}
D.~Sullivan, D.~Allan, D.~Howe, and F.~Walls, ``Characterization of clocks and
  oscillators,'' {NIST} tech. note 1337, NIST, U.S Department of Commerce,
  National Institute of Standards and Technology, 1990.
\newblock Online available at
  \url{https://nvlpubs.nist.gov/nistpubs/Legacy/TN/nbstechnicalnote1337.pdf}.

\bibitem{gum95}
``{Guide to the Expression of Uncertainty in Measurement}.'' ISO/TAG 4.
  Published by ISO, 1993 (corrected and reprinted, 1995) in the name of the
  BIPM, IEC, IFCC, ISO, UPAC, IUPAP and OIML, 1995.
\newblock {ISBN} number: 92-67-10188-9, 1995.

\bibitem{Zhang2006}
N.~F. Zhang, ``{Calculation of the uncertainty of the mean of autocorrelated
  measurements},'' {\em Metrologia}, vol.~43, no.~4, pp.~S276--S281, 2006.

\bibitem{Zieba2010}
A.~Zi{\k{e}}ba, ``{Effective number of observations and unbiased estimators of
  variance for autocorrelated data-an overview},'' {\em Metrology and
  Measurement Systems}, vol.~17, no.~1, pp.~3--16, 2010.

\bibitem{Zieba2011}
A.~Zi{\k{e}}ba and P.~Ramza, ``{Standard deviation of the mean of
  autocorrelated observations estimated with the use of the autocorrelation
  function estimated from the data},'' {\em Metrology and Measurement Systems},
  vol.~18, no.~4, pp.~529--542, 2011.

\bibitem{Prillaman2010}
J.~Prillaman, E.~Powers, B.~Fonville, S.~Mitchell, and E.~Goldberg,
  ``{Continued Evaluation of Carrier-Phase GNSS Timing Receivers for UTC/TAI
  Applications},'' tech. rep., NAVAL OBSERVATORY WASHINGTON DC, 2010.

\bibitem{Ray2003}
J.~Ray and K.~Senior, ``{IGS/BIPM pilot project: GPS carrier phase for
  time/frequency transfer and timescale formation},'' {\em Metrologia},
  vol.~40, no.~3, pp.~S270--S288, 2003.

\bibitem{Weinbach2013a}
U.~Weinbach, {\em {Feasibility and impact of receiver clock modeling in precise
  GPS data analysis}}.
\newblock PhD thesis, Leibniz University Hannover, 2013.

\bibitem{CircT}
{BIPM Circular T}. \url{http://www.bipm.org/jsp/en/TimeFtp.jsp}.

\bibitem{Benkler2015}
E.~Benkler, C.~Lisdat, and U.~Sterr, ``{On the relation between uncertainties
  of weighted frequency averages and the various types of Allan deviations},''
  {\em Metrologia}, vol.~52, no.~4, pp.~565--574, 2015.

\bibitem{Riehle2018}
F.~Riehle, P.~Gill, F.~Arias, and L.~Robertsson, ``{The CIPM list of
  recommended frequency standard values: guidelines and procedures},'' {\em
  Metrologia}, vol.~55, no.~2, pp.~188--200, 2018.

\bibitem{Jiang_2018}
Z.~Jiang, V.~Zhang, Y.-J. Huang, J.~Achkar, D.~Piester, S.-Y. Lin, W.~Wu,
  A.~Naumov, S.~hoon Yang, J.~Nawrocki, I.~Sesia, C.~Schlunegger, Z.~Yang,
  M.~Fujieda, A.~Czubla, H.~Esteban, C.~Rieck, and P.~Whibberley, ``Use of
  software-defined radio receivers in two-way satellite time and frequency
  transfers for {UTC} computation,'' {\em Metrologia}, vol.~55, no.~5,
  pp.~685--698, 2018.

\bibitem{Jiang_2019}
Z.~Jiang, V.~Zhang, T.~E. Parker, G.~Petit, Y.-J. Huang, D.~Piester, and
  J.~Achkar, ``Improving two-way satellite time and frequency transfer with
  redundant links for {UTC} generation,'' {\em Metrologia}, vol.~56, no.~2,
  p.~025005, 2019.

\bibitem{Percival1993}
D.~B. Percival, ``{Three curious properties of the sample variance and
  autocovariance for stationary processes with unknown mean},'' {\em The
  American Statistician}, vol.~47, no.~4, pp.~274--276, 1993.

\end{thebibliography}

\end{document}